\documentclass[submission, Phys]{SciPost}
\usepackage{color,amsmath,amssymb,multirow,hyperref,booktabs,graphicx,mathtools,url} 

\graphicspath{{./figs/}}
\newcommand{\arxiv}[1]{\href{http://arxiv.org/abs/#1}{arXiv:#1}}
\newcommand{\urlx}[1]{\href{#1}{#1}}

\newcommand\one{\leavevmode\hbox{\small1\normalsize\kern-.33em1}}

\newcommand{\met}{\slashchar{E}_T}

\newcommand{\gev}{\text{GeV}}

\def\slashchar#1{\setbox0=\hbox{$#1$}           % set a box for #1
   \dimen0=\wd0                                 % and get its size
   \setbox1=\hbox{/} \dimen1=\wd1               % get size of /
   \ifdim\dimen0>\dimen1                        % #1 is bigger
      \rlap{\hbox to \dimen0{\hfil/\hfil}}      % so center / in box
      #1                                        % and print #1
   \else                                        % / is bigger
      \rlap{\hbox to \dimen1{\hfil$#1$\hfil}}   % so center #1
      /                                         % and print /
   \fi}

\newcommand{\eg}{\textsl{e.g.}\;}
\newcommand{\ie}{\textsl{i.e.}\;}

\newcommand{\be}{\begin{eqnarray*}}
\newcommand{\ee}{\end{eqnarray*}}

\newcommand{\bee}{\begin{eqnarray}}
\newcommand{\eee}{\end{eqnarray}}
\newcommand{\beeq}{\begin{equation}}
\newcommand{\eeeq}{\end{equation}}

\begin{document}

\begin{center}{\Large \textbf{
QCD or What?
}}\end{center}

\begin{center}
Theo Heimel\textsuperscript{1},
Gregor Kasieczka\textsuperscript{2},
Tilman Plehn\textsuperscript{1}, and
Jennifer M Thompson\textsuperscript{1}
\end{center}

\begin{center}
{\bf 1} Institut f\"ur Theoretische Physik, Universit\"at Heidelberg, Germany\\
{\bf 2} Institut f\"ur Experimentalphysik, Universit\"at Hamburg, Germany
plehn@uni-heidelberg.de
\end{center}

\begin{center}
\today
\end{center}

% For convenience during refereeing: line numbers
%\linenumbers

\section*{Abstract}
{\bf Autoencoder networks, trained only on QCD jets, can be used to
  search for anomalies in jet-substructure. We show how, based either on
  images or on 4-vectors, they identify jets from decays of arbitrary
  heavy resonances.  To control the backgrounds and the underlying
  systematics we can de-correlate the jet mass using an adversarial
  network. Such an adversarial autoencoder allows for a general and at
  the same time easily controllable search for new physics.  Ideally,
  it can be trained and applied to data in the same phase space
  region, allowing us to efficiently search for new physics using
  un-supervised learning.}

\vspace{10pt}
\noindent\rule{\textwidth}{1pt}
\tableofcontents\thispagestyle{fancy}
\noindent\rule{\textwidth}{1pt}
\vspace{10pt}

\newpage
%%%%%%%%%%%%%%%%%%%%%%%%%%%%%%%%%%%%%%%%%%%%%%%%%%%%%%%%%%%%%%%%%%%%%%
\section{Introduction}
\label{sec:intro}

Since the start of the LHC, jets have turned from an experimental
annoyance to the most interesting and powerful analysis objects.
Together with a vastly improved understanding of multi-jet kinematics,
we have learned how to use jet constituents to identify LHC
signals~\cite{bdrs}. This way, jets no longer serve as universal
analysis objects, but merely separate jet-level observables from
subjet observables~\cite{early_top,heptop1}. A second development in
LHC analyses is that we compare simulated and observed jet events at
the detector level, instead of first-principles theory and
data~\cite{template,heptop3,shower_deco}. This raises the question why
we still rely on intermediate high-level observables rather than
low-level observables like 4-momenta of particle flow objects. The
latter are driving deep learning applications at the LHC, where the
term deep learning really describes the shift from high-level to
low-level input
observables~\cite{early_images,momenta,aussies,gan,deeptop,david,lola,quark_gluon}.
In this paper we show how this new approach allows us to tackle the
basic question:
\begin{center} 
\textsl{Do our observed jets really look like QCD jets?}
\end{center} 

In addition to testing well-defined hypotheses, neural networks also
allow us to search for anomalies without ever defining a signal. We
can just study QCD jets in data and use machine learning techniques to search
for non-QCD patterns. The appropriate network architecture are
autoencoders~\cite{auto_old,auto_sebastian,ml_review}, networks which
compress their information to search for patterns which are no longer
described by the compressed representation.\medskip

Once we abandon high-level subjet observables we can choose our input
format to deep learning analysis tools. This allows us to pick a data 
format that is best suited to a given problem. The most frequently used
format are jet images, calorimeter entries in the azimuthal angle vs
rapidity plane, analyzed through image recognition~\cite{cnn}.  They
can be used to identify hadronic decays of weak
bosons~\cite{early_images,aussies,gan} or top
quarks~\cite{deeptop,david}, or to distinguish quark-like from
gluon-like jets~\cite{quark_gluon}. A limiting factor for jet images
are measurements with vastly different angular resolution, like
calorimeter and tracker measurement~\cite{quark_gluon,lola}. For
example in this case we can directly use 4-momenta~\cite{momenta} and,
inspired by graph convolutional networks, analyze them efficiently
using the Minkowski metric~\cite{lola}. These approaches can also be
generalized to search for new physics at the event
level~\cite{new_physics}. For all network setups we can visualize their
behavior based on truth-level information in Monte Carlo
simulations~\cite{quark_gluon,deeptop,what_learning}. \medskip

Deep learning applications to jet physics at the LHC face three key limitations:
\begin{enumerate}
\setlength\itemsep{-0.3em}
\item the availability of training data;
\item systematic uncertainties in our understanding of the training data;
\item control over the exact physics question which the network
  answers.
\end{enumerate}
While most available studies answer well-defined physics questions
based on labelled or simulation data, the first limitation can be
tackled by switching to weakly supervised
learning~\cite{weakly_supervised}. In this paper we will go even
further and show how autoencoders work in the absence of a signal
sample. The second challenge can for example be addressed with adversarial
networks, de-correlating for example kinematic information or theory
assumptions~\cite{ddt,pivot,adversarial1,adversarial2,adversarial_auto}. 
Alternatively, refiner networks~\cite{refiner} can be used to improve the quality of simulation.
For our
autoencoder we will de-correlate kinematic information like the jet
mass, generating experimental control regions and controlling
systematics related to the composition of the training sample. This
way, the autoencoder can be trained and applied on the same phase
space and avoid systematic uncertainties from relating simulation to
data or background to signal phase space regions.  The same de-correlation
technique also allows us to tackle the third challenge. A promising
approach in this direction is to combine well-understood features like
mass peaks with implicit, orthogonal information~\cite{cwola}. More
generally, we will show how any well-defined physics effect can be
de-correlated from the autoencoder analysis, allowing us to construct
control regions, side bands, or a flat or smoothly falling spectrum suitable for a bump
hunt in any observable needed for a given analysis.\medskip

In Sec.\ref{sec:auto_cnn} we will start by constructing an
autoencoder~\cite{auto_old,auto_sebastian,ml_review} based on a
convolutional network~\cite{auto_chollet}, in our case the image-based
\textsc{DeepTop} tagger~\cite{deeptop,david}. Alternatively, we can
analyze 4-vectors in an autoencoder version of the
\textsc{DeepTopLoLa} tagger, as shown in Sec.~\ref{sec:auto_lola}.
Next, we will control what kind of information the network uses by
taking out the jet mass distribution through an adversarial network in
Sec.~\ref{sec:adv}.  This allows us to devise a convincing side band
analysis on the jet mass for the anomaly search~\cite{cwola}. With the
help of these sidebands we can study the stability of the autoencoder
network trained on not fully controlled, impure QCD samples in
Sec.~\ref{sec:impure}. After establishing our new methods using top
tagging we will test them on scalar decays to four jets in
Sec.~\ref{sec:scalar} and on non-QCD showers in Sec.~\ref{sec:shower}.

%%%%%%%%%%%%%%%%%%%%%%%%%%%%%%%%%%%%%%%%%%%%%%%%%%%%%%%%%%%%%%%%%%%%%%
\section{Autoencoded QCD vs tops} 
\label{sec:auto}

The aim of our study is to identify jets with an exotic, non-QCD
origin using a neural network that is only trained on QCD jets.  This
can be done with autoencoder networks, which are stacks of networks layers with
an intermediate set of bottleneck layers with a strongly reduced
number of units, corresponding to a latent space with reduced
dimensionality. Such a bottleneck can be added to convolutional
networks~\cite{auto_chollet}, but it can also be added to a
\textsc{LoLa}-like network working on constituent 4-vectors. The main
structural change is that autoencoders do not work towards an output
value which, assuming the right loss function, gives a probability for
a jet being either signal or background. Instead, the network on both
sides of the bottleneck is approximately symmetric, and the loss
function is the difference between the input and the output. Once we
run such a trained network on a test sample the loss function will
tell us how well the network with its bottleneck encodes the
features of the test sample.

An established, albeit non-glamorous benchmark for subjet studies are
boosted hadronic top decays. This is why we first test our new
autoencoder setup, trained on QCD jets, for anomalous top
jets. After we benchmark autoencoders for image-based and
4-vector-based architectures, we will introduce a de-correlation with
the jet mass. This approach can be immediately generalized to any
other variable, defining plenty of control regions and side bands to
control the autoencoder in an experimental reality.

%%%%%%%%%%%%%%%%%%%%%%%%%%%%%%%%%%%%%%%%%%%%%%%%%%%%%%%%%%%%%%%%%%%%%%
\subsection{Jet images} 
\label{sec:auto_cnn}

As long as we focus on the pixellated energy, we can analyze
jets using a convolutional neural network (CNN) to learn jet images. Our autoencoder architecture is
based on the \textsc{DeepTop} tagger~\cite{deeptop}, with 
significant improvements especially to the image pre-processing,
developed in Ref.~\cite{david}. 

Our top and QCD samples are similar to the sample used in our
\textsc{DeepTop} studies~\cite{deeptop,lola}. We simulate top pair and
di-jet production with \textsc{Pythia8.2.15}~\cite{pythia} and
\textsc{Delphes3.3.2}~\cite{delphes} for a collider energy of 14~TeV.
For the QCD sample we do not distinguish between hard quarks and
gluons.  
%We ignore multi-parton interaction and pile-up, assuming that
%there are specialized tools available to remove it.
While the simulation used for these studies did not include effects of multi-parton interaction and pile-up, this is not a fundamental limitation of the proposed approach. Autoencoders can also be applied to the constituents of a jet after applying standard experimental techniques for the removal of pile-up~\cite{pileup}. A combination with grooming algorithms is possible as well, but would potentially limit the sensitivity as grooming makes explicit assumptions on how a shower ought to unfold.

Similarly, no detailed detector simulation was included. We expect the autoencoder to learn novel jet-shape variables from the distributions of constituents. There is no a-priori reason why these jet shapes would suffer from larger effects due to the detector simulation than widely used variables like groomed mass, n-subjettiness or energy correlation functions. For the practical application of the autoencoder we foresee training on data, making this technique even less subject to differences between data and simulation than ordinary approaches.

\begin{figure}[t]
\includegraphics[width=0.46\textwidth]{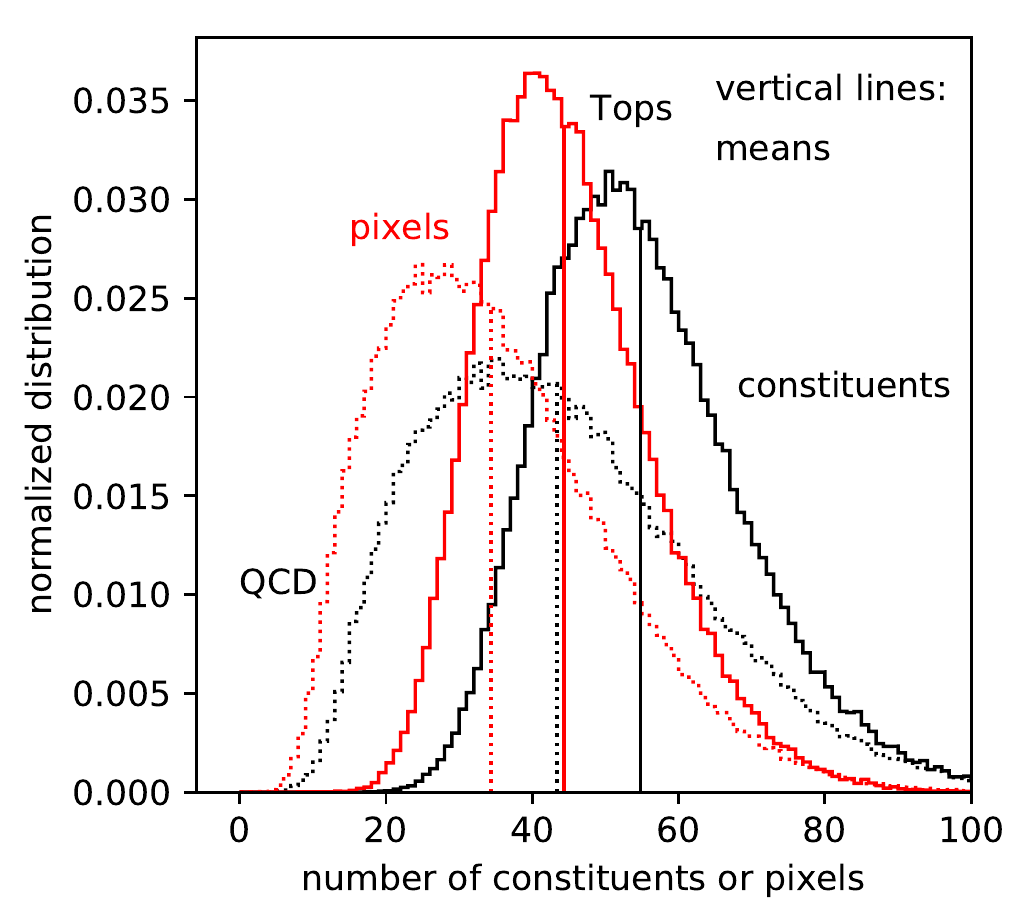}
\hspace*{0.03\textwidth}
\includegraphics[width=0.51\textwidth]{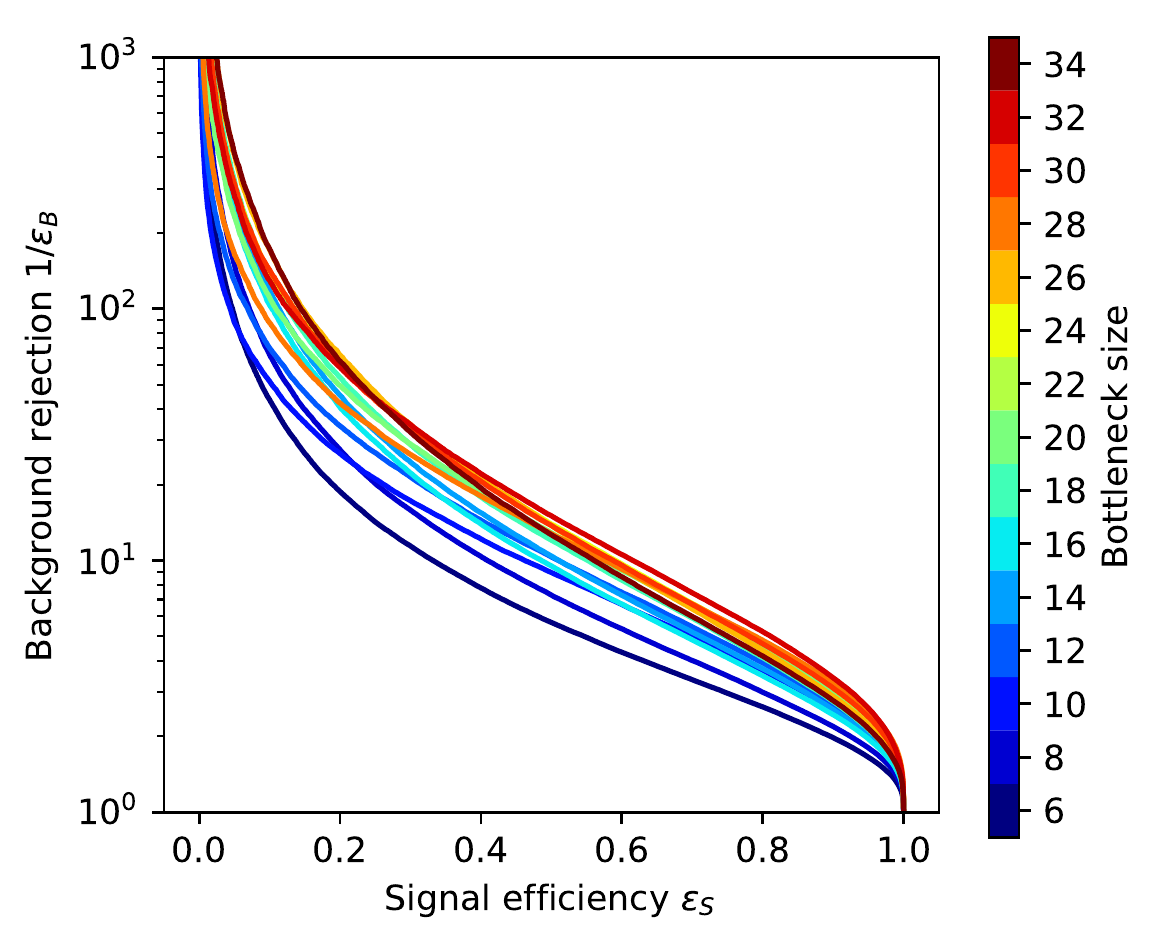}
\caption{Left: numbers of constituents and of
  non-zero pixels for tops and QCD jets, 400,000 jets in
  total. Right: ROC curves for the image-based autoencoder identifying
  anomalous top jets for different bottleneck sizes.}
\label{fig:roc_cnn}
\end{figure}

 The
substructure containers are fat anti-$k_T$ jets~\cite{anti_kt} with
distance parameter $R=0.8$, defined by \textsc{FastJet3.1.3}~\cite{fastjet}. 
They are required to have a transverse momentum in the range
\begin{align} 
p_{T,j} = 550~...~650~\gev \; .
\end{align}
In addition they must be central, $|\eta_j| < 2$. For all signal
jets we require both the truth-level partonic top and its decay products to be within the area of
the fat jet. 
 The inputs of the subjet analysis are particle flow
objects~\cite{particle_flow} from the \textsc{Delphes} E-flow. In the
left panel of Fig.~\ref{fig:roc_cnn} we show the number of particle
flow constituents for signal and background jets. The main feature
is that already based on the larger number of constituents we could
identify the hadronic top decays.\medskip

%------------------------------------------------
%------------------------------------------------

Following Ref.~\cite{david} we employ an improved pre-processing of
the jet images, most notably applied before pixelization. This
approach is directly motivated by the particle flow approach, which
combines the coarse calorimeter information with the high-resolution
tracker and provides us with a set of high-resolution 4-vectors. The
center of the image is not defined by the hardest object, but by the
$k_T$-weighted centroid of the fat jet constituents. The major
principle axis is then turned to 12 o'clock. Finally, the image is
flipped along the $x$-axis and $y$-axis, to ensure that the hardest
constituent is located in the upper right quadrant. Only after this
pre-processing we pixelize the images into a $40 \times 40$-pixel
image, covering $\eta = -0.57~...~0.57$ and $\phi = -0.69~...~0.69$
around the center of the fat jet.  The entries of the calorimeter
images are given by the transverse momentum entering the detector
cell, \ie the sum of the transverse momenta of all particle flow
objects covered by a pixel.  Also in the left panel of
Fig.~\ref{fig:roc_cnn} we show the number of non-zero pixels per
image. The full image with its 1600 pixels is indeed sparsely filled.
Each of the pixels is finally normalized to the sum of all pixels in
the jet image.  These images define the input and the output format of
the autoencoder network.\medskip

%------------------------------------------------
\begin{figure}[b!]
\centering
\includegraphics[width=1.0\textwidth]{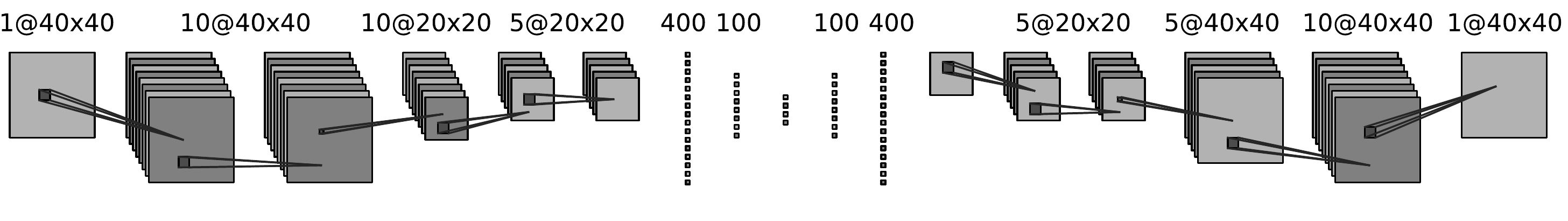}
\vspace*{-6mm}
\caption{Architecture of the image-based autoencoder network. The $40
  \times 40$ images are average-pooled to $20 \times 20$ images before
  entering the bottleneck. The dense units are first reduced from 400
  to 100, the minimum size at the bottleneck is variable.}
\label{fig:arch_cnn}
\end{figure}
%------------------------------------------------

The architecture of the autoencoder network is shown in
Fig.~\ref{fig:arch_cnn}. We use \textsc{Keras}~\cite{keras} combined
with \textsc{Tensorflow}~\cite{tensorflow} to build our network.  It
is almost symmetric between the input and the output. The loss
function is simply
\begin{align}
L_\text{auto} = \sum_\text{1600 pixels} \left( k_T^\text{norm,in} - k_T^\text{auto} \right)^2 \; ,
\label{eq:loss_cnn}
\end{align}
in terms of the normalized input image and the autoencoder output
image. We use the PReLU activation function throughout the network, to
avoid a zero pseudo-solution, except for a linear activation function
in the last layer. We use the \textsc{Adam} optimizer~\cite{adam} for
training the network.
%the learning rate.

The autoencoder is trained on 100,000 QCD or background jets for up to
100 epochs and allow for an early stopping after ten epochs with
stable loss. Our test sample consists of 200,000 top jets and 200,000
QCD jets. The large test sample allows for a study of the performance
on several independent samples, confirming that our ROC curves are
stable.  For a variable cut in the loss function we can evaluate the
composition of the signal-like jets in terms of true top and true QCD
jets. These two fractions define a ROC curve, as shown in the right
panel of Fig.~\ref{fig:roc_cnn}.  For these curves we vary the size of
the bottleneck from 6 to 34 units in the smallest dense layer shown in
Fig~\ref{fig:arch_cnn}. We see a sizeable variation with the
bottleneck size, developing a stable high-performance plateau between
20 and 34.  It gives a stable area under curve (AUC) around 0.89 with
a loss around $10^{-5}$ per pixel.  The size of the bottleneck has to
be compared with the initially 1600 pixels, of which 10 to 70 are
non-zero, and which the CNN pools to 400 combined pixels.  This large
bottleneck size indicate that the image architecture is not perfectly
adapted to encode the relevant QCD vs tops information in a small
network layer.

The large size of the test sample allows us to evaluate our
autoencoder on separate, statistically independent test samples. While
the corresponding spread does not account for systematics
uncertainties related to the training, especially the training on
data, it defines a statistical uncertainty of the autoencoder.  It is
shown as widths of the ROC curves, which are generated by evaluating
the network on ten independent test samples with 20,000 QCD jets and
20,000 top jets each.

%%%%%%%%%%%%%%%%%%%%%%%%%%%%%%%%%%%%%%%%%%%%%%%%%%%%%%%%%%%%%%%%%%%%%%
\subsection{LoLa} 
\label{sec:auto_lola}

%------------------------------------------------
\begin{figure}[b!]
\centering
\includegraphics[width=0.7\textwidth]{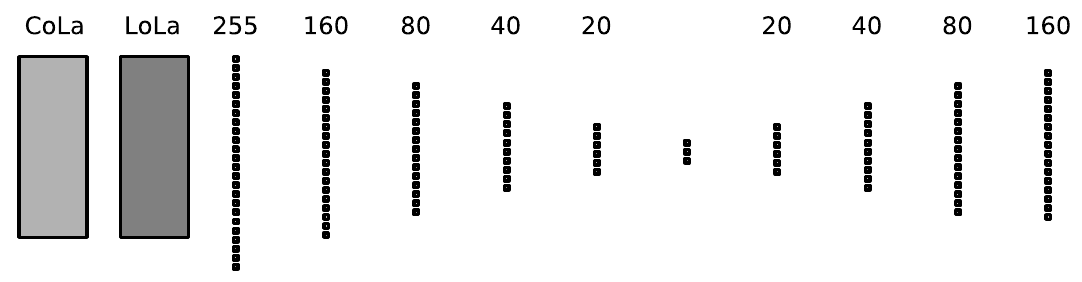}
\vspace*{-6mm}
\caption{Architecture of the 4-vector-based autoencoder network. The
  255 input units correspond to 55 \textsc{LoLa}-vectors with $4+1$
  entries each. The output only consists of 160 units, because the
  extended 4-vectors only carry four independent observables.}
\label{fig:arch_lola}
\end{figure}
%------------------------------------------------

When we want to include information beyond the calorimeter output, we can for
example use the neural network based on the constituent 4-vectors
developed for the \textsc{DeepTopLoLa} tagger~\cite{lola}. It starts
from a set of measured 4-vectors sorted by transverse momentum
\begin{align}
( k_{\mu,i} ) = 
\begin{pmatrix}
k_{0,1} &  k_{0,2} & \cdots &  k_{0,N} \\
k_{1,1} &  k_{1,2} & \cdots &  k_{1,N} \\ 
k_{2,1} &  k_{2,2} & \cdots &  k_{2,N} \\ 
k_{3,1} &  k_{3,2} & \cdots &  k_{3,N}  
\end{pmatrix} \; .
\label{eq:def_input}
\end{align}
Following the left panel of Fig.~\ref{fig:roc_cnn} we use $N=40$
constituents, after checking that an increase to $N=120$ does not make a
measurable difference. For jets with fewer constituents we naturally
fill the entries remaining in the soft regime with zeros.  

%------------------------------------------------
\begin{figure}[t]
\includegraphics[width=0.50\textwidth]{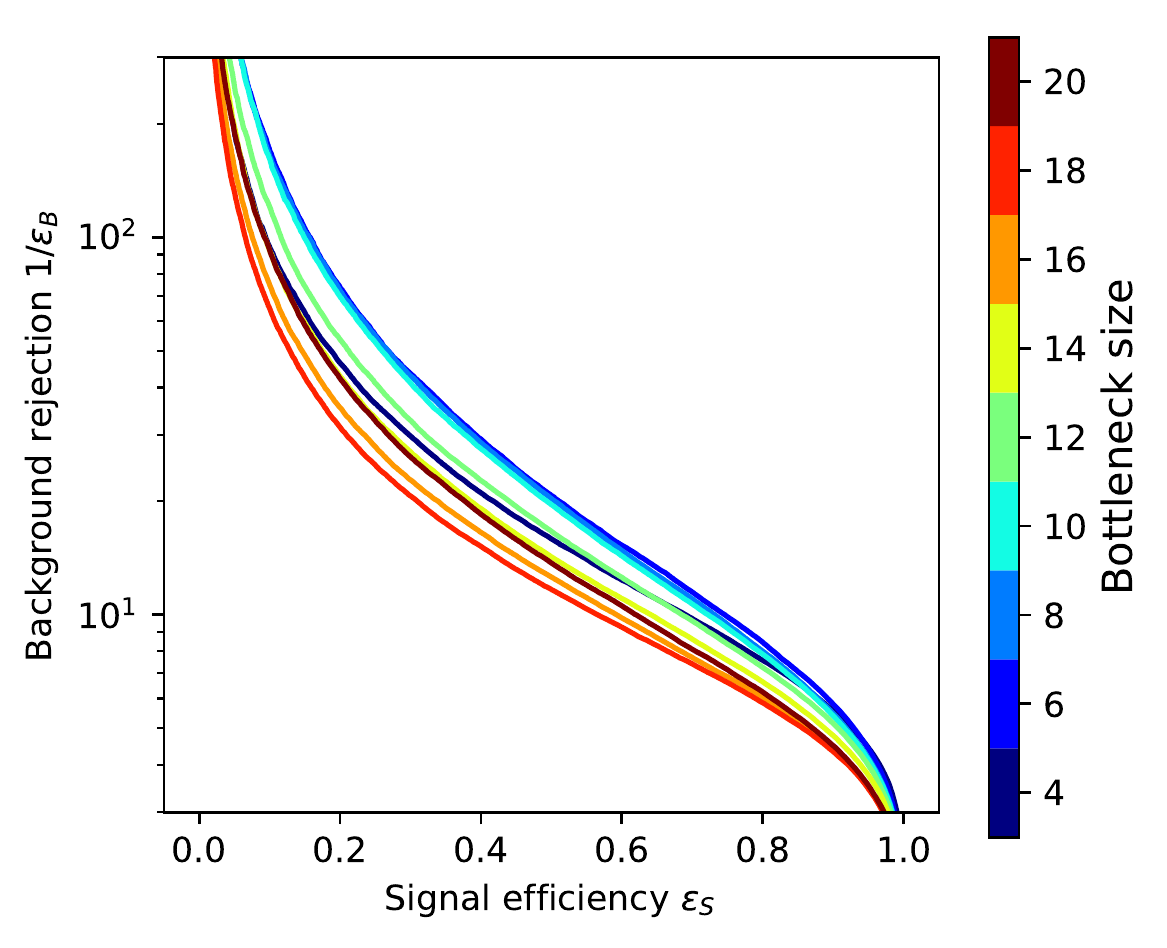}
\hspace*{0.05\textwidth}
\includegraphics[width=0.42\textwidth]{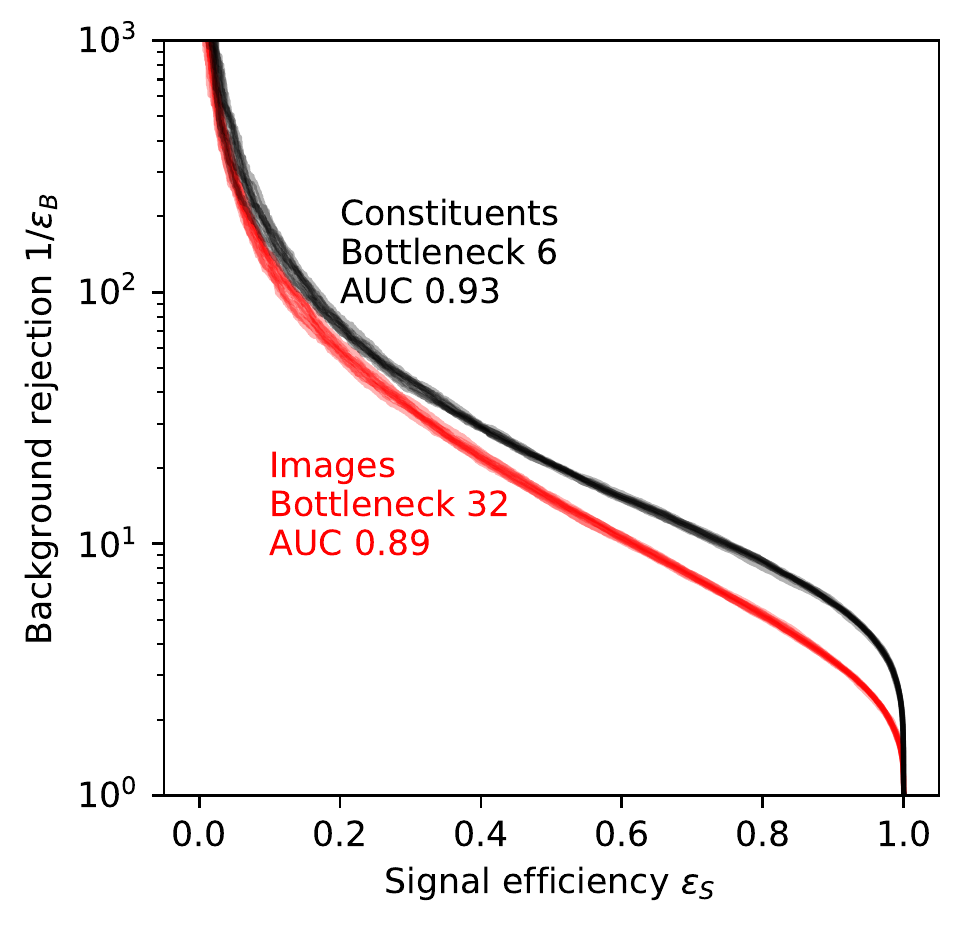}
\caption{Left: ROC curves for the 4-vector-based or \textsc{LoLa}
  autoencoder identifying anomalous top jets for different bottleneck
  sizes. Right: comparison of the ROC curves for the image-based and
  the 4-vector-based autoencoders. The widths of the lines show the
  variation based on ten independent test samples for fixed training.}
\label{fig:roc_lola}
\end{figure}
%------------------------------------------------

To remove all information from the jet-level kinematics we boost all
4-momenta into the rest frame of the fat jet. This also improves the
performance of our network.  Inspired by recombination jet
algorithms we can add linear combinations of these 4-vectors with a
trainable matrix $C_{ij}$, defining a combination layer
\begin{align}
k_{\mu,i} \stackrel{\text{CoLa}}{\longrightarrow}
\widetilde{k}_{\mu,j} 
= k_{\mu,i} \; C_{ij} 
\qquad \text{with} \quad
C = 
 \begin{pmatrix}
1 & 1 & 0 & \cdots & 0      & C_{1,N+2} & \cdots & C_{1,M} \\[-2mm]
\vdots & 0 & 1 &  & \vdots & C_{2,N+2} & \cdots & C_{2,M} \\[-2mm]
\vdots & \vdots & \vdots & \ddots & 0 & \vdots &  & \vdots \\
1 & 0 & 0 & \cdots & 1 &C_{N,N+2} & \cdots & C_{N,M} 
 \end{pmatrix} \; .
\label{eq:cola}
\end{align}
We allow for $M=10$ trainable linear combinations.  These combined
4-vectors carry information on the hadronically decaying massive
particles. In the original \textsc{LoLa} approach we map the momenta
$\tilde{k}_j$ onto observable Lorentz scalars and related
observables~\cite{lola}. Because this mapping is not easily invertible
we do not use it for the autoencoder.  Instead, we extend the
4-vectors by another component containing the invariant mass,
\begin{align}
\tilde{k}_j 
= \begin{pmatrix*}[c]
  \tilde{k}_{0,j} \\ \tilde{k}_{1,j} \\ \tilde{k}_{2,j} \\ \tilde{k}_{3,j}
  \end{pmatrix*}
\stackrel{\text{LoLa}}{\longrightarrow}
  \begin{pmatrix*}[c]
  \tilde{k}_{0,j} \\ \tilde{k}_{1,j} \\  \tilde{k}_{2,j} \\ \tilde{k}_{3,j} \\ \sqrt{\tilde{k}_j^2}
  \end{pmatrix*} \; .
\label{eq:lola}
\end{align}
This defines a set of 51 extended 4-vectors, which form the input to
our neural network. Again, we use \textsc{Keras}~\cite{keras} combined
with \textsc{Tensorflow}~\cite{tensorflow}. Its architecture is shown
in Fig.~\ref{fig:arch_lola}. The layer immediately after the
\textsc{LoLa} contains $51 \times (4+1) = 255$ units. Between the
second layer after \textsc{LoLa} and the last layer, the autoencoder
network is symmetric. The final output consist of 40 4-vector-like
objects, which can be compared with the corresponding second
layer. The loss function is
\begin{align}
L_\text{auto} = 
\sum_{j=1}^{40} 
\sum_{i=0}^3 
\left( \tilde{k}_{i,j}^\text{in} - \tilde{k}_{i,j}^\text{auto} \right)^2 \; .
\label{eq:loss_lola}
\end{align}
As for the images we use the PReLU activation function, except for the
last layer with its linear activation function, and the \textsc{Adam}
optimizer for the learning rate~\cite{adam}.\medskip

In the left panel of Fig.~\ref{fig:roc_lola} we show the ROC curves
for the 4-vector-based tagger for different choices of the bottleneck
size. We now find the best result for a very small bottleneck with at
most 10 units. The stable AUC value is around 0.92 with a loss around
$10^{-5}$ per pixel. Such small functional bottlenecks reflect the
fact that with the \textsc{CoLa/LoLa} structure we have encoded a lot
of the relevant information in appropriate physics terms~\cite{lola}.

Finally, in the right panel of Fig.~\ref{fig:roc_lola} we compare the
best-performing image-based and 4-momentum-based autoencoders. The
widths of the lines are again generated by evaluating the network on
ten independent test samples.  The main feature in this plot is that
the \textsc{LoLa}-autoencoder does better than the image-based
autoencoder. This is a result of the smaller possible bottleneck size,
because the \textsc{LoLa} architecture is optimized to extract the
leading discriminating features most efficiently. While this gives an
advantage to the pure autoencoder, we will see the other side of the
same medal in the next section.

%%%%%%%%%%%%%%%%%%%%%%%%%%%%%%%%%%%%%%%%%%%%%%%%%%%%%%%%%%%%%%%%%%%%%%
\subsection{De-correlating the mass} 
\label{sec:adv}

Neural networks separating signal and background jets after fully
supervised training on labelled data are, in theory, straightforward
to calibrate and understand. The problem at the LHC is that we hardly
ever have enough labelled data to train such networks for relevant new
physics searches --- especially when the goal is to tag new resonances. 
Our autoencoder responds to this problem by limiting
the training to QCD jets only and by only asking if a given
data set is described well by QCD or any other standard assumption. On
the other hand, the more weakly the question is defined,
the more important it is to control what the neural network actually
learns. This is especially true when we use the network on low-level
information rather then established high-level kinematic
observables~\cite{non_nn}.\medskip 

An established way to test a network is to exclude known, well-defined
pieces of information from it through adversarial
networks~\cite{pivot,adversarial1,adversarial2,adversarial_auto}. They consist of two
networks playing against each other. Similar to generative adversarial
networks, they can be used to train a network as an equivalent
replacement for another data generator. In our application the
additional adversary is trained to extract for instance the jet mass
from the autoencoder output described in Eq.\eqref{eq:loss_cnn}. In
this image-based case a naive adversary loss function would read
\begin{align}
L_\text{adv}(M) = 
\left[ \widetilde{M} \left( \left| k_{T,i}^\text{adv} - k_{T,i}^\text{auto} \right| \right) - M \right]^2 \; ,
\end{align}
with the inputs $k_{T,i}^\text{auto}$, the outputs
$k_{T,i}^\text{adv}$, the given jet mass $M$, and the trained proxy to
the jet mass $\widetilde{M}$. As we will discuss below, for our study
we replace the exact function $\widetilde{M}$ with a binned determination of the
jet mass~\cite{adversarial1}.  The combined loss function which
replaces Eq.\eqref{eq:loss_cnn} for the autoencoder can be written in
terms of a Lagrangian multiplier~\cite{adversarial1,pivot}
\begin{align}
L = L_\text{auto} - \lambda \, L_\text{adv}(M) \; .
\label{eq:loss_adv}
\end{align}
The Lagrangian multiplier $\lambda$ introduces a boundary condition,
$L_\text{adv} \to 0$, in case the adversary learns the mass perfectly.
The value of $\lambda$ determines the balance between the two
networks. While the task of the autoencoder network is to describe the
QCD training data, the adversary extracts the jet mass from the
autoencoder output. Playing against each other and minimizing the
combined loss function with the relative sign, the combined network
wants the adversary to be as unsuccessful as possible. The adversarial
autoencoder will hence avoid all information on the jet mass or any
other boundary condition. Note that at least for the top jets this
only affects the fat jet mass and still leaves us with the $W$-mass in
the clustering history.\medskip

As a starting point, we show the jet mass distribution after applying
the image-based autoencoder. We know from many studies that the jet
mass is a powerful observable in separating QCD jets
from hadronically decaying heavy states. On the other hand, since we also
know that a small fraction of QCD jets will feature large jet masses,
we expect to see a top signal as a jet mass peak over a smooth QCD jet
background.  

\begin{figure}[t]
\includegraphics[width=0.45\textwidth]{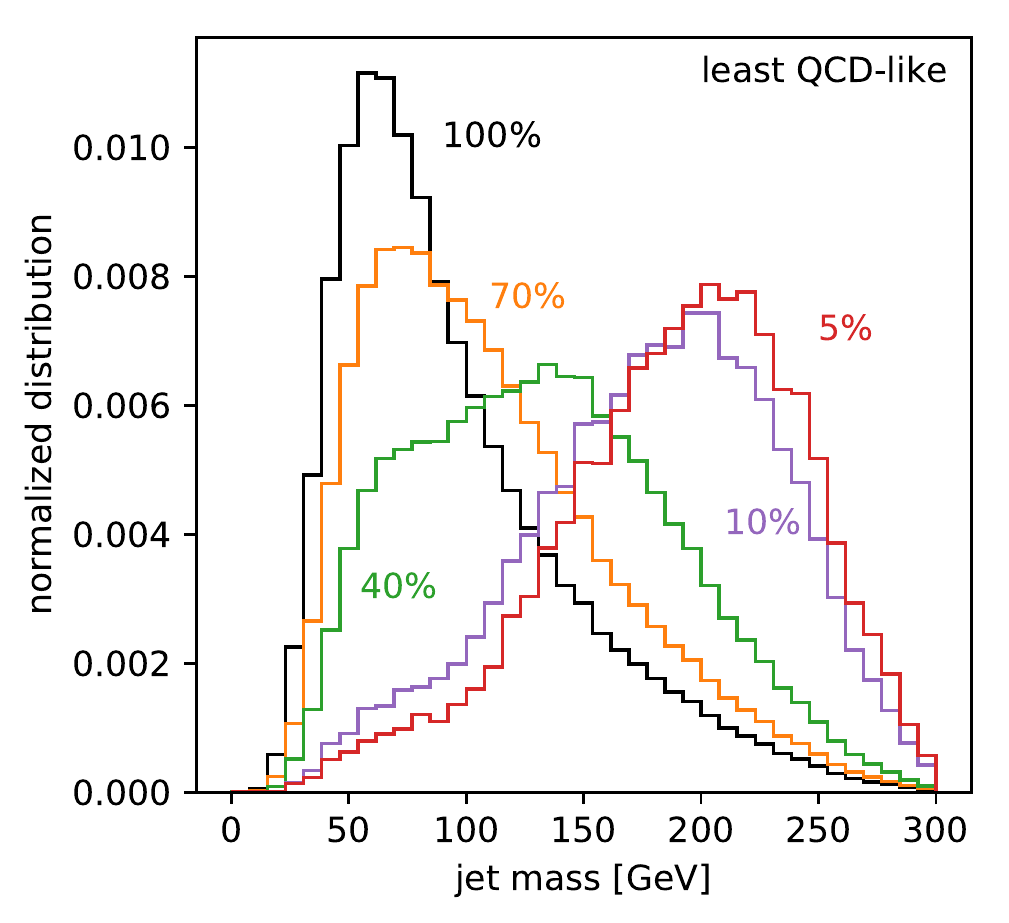}
\hspace*{0.03\textwidth}
\includegraphics[width=0.45\textwidth]{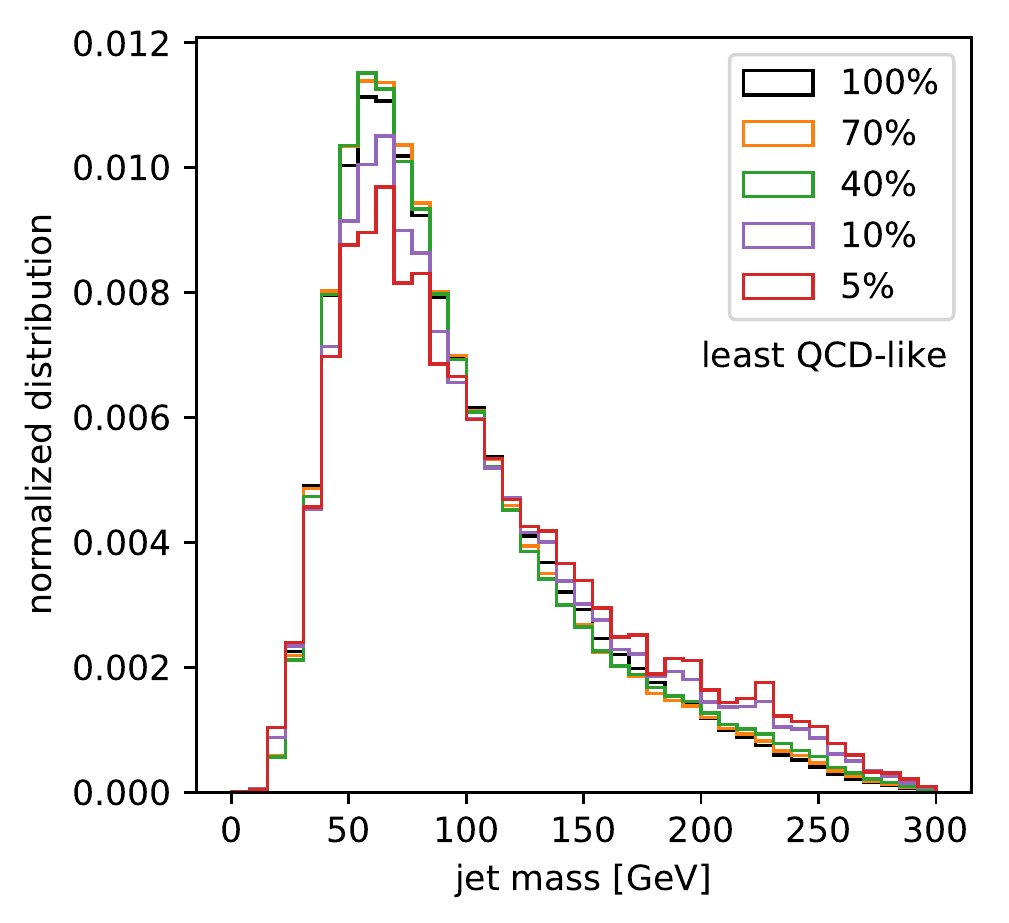}
\caption{Left: jet mass distributions from the image-based autoencoder
  applied to QCD jets. The different lines show the full sample up to
  the 5\% least QCD-like jets, defined by the autoencoder loss
  function. Right: the same jet mass distributions, but for the
  QCD-trained adversarial autoencoder network.}
\label{fig:adv_mass}
\end{figure}

In the left panel of Fig.~\ref{fig:adv_mass} we show jet mass
distributions for QCD jets in slices of the autoencoder loss function.
The per-centile ranges from all QCD jets to the 5\% least QCD-like of
all QCD jets. For the full jet sample we see the expected peak at
small $m_j \approx 50$~GeV with a long tail extending beyond
300~GeV. For the least QCD-like jets in the pure QCD sample a peak at
$m_j \approx 200$~GeV appears.  This means that the cut on the
autoencoder output badly shapes the background and makes it
signal-like. This defines the task of the adversarial network: provide
a smooth jet mass distribution for QCD jets, independent of the value
of the autoencoder loss function; or in other words, de-correlate the
jet mass from the autoencoder.\medskip

%------------------------------------------------
%------------------------------------------------

Again, we use \textsc{Keras}~\cite{keras} and
\textsc{Tensorflow}~\cite{tensorflow} with the
\textsc{Adam}~\cite{adam} optimizer for the combined adversarial
network. The image-based autoencoder part of the network is described
in Fig.~\ref{fig:arch_cnn}; the adversarial part consists of eight
dense layers with 800, 400, 200, 100, 50, 25, 10, and 12 units. We now
train this network on 600,000 QCD jets.  The output layer corresponds
to 10 pre-defined slices in the jet mass, binned such that they are
populated by the same number of QCD jets. On each side we add overflow
bins which are not populated by QCD jets. The task of the adversary is
not to extract the exact jet mass value, but to determine the
probabilities for the jet mass to fall into each bin. This statistical
interpretation requires a multi-label cross entropy as the adversary
loss function~\cite{adversarial1}.  All layers use the ReLU activation
function except for the last layer, where a SoftMax activation
function guarantees that all 12 probabilities sum to one.  When
training on the combined loss function, each epoch is split into
batches of size 128. For each batch we first train the autoencoder
using the combined loss function of Eq.\eqref{eq:loss_adv} and then
train the adversary with only the adversary loss function. The size of
the Lagrangian multiplier is chosen such that the two contributions to
the loss function are of similar size, \ie it balances the
de-correlation vs the discrimination power of the network. For
instance, the jet mass distribution for $\lambda = 5 \cdot 10^{-4}$,
shown in the right panel of Fig.~\ref{fig:adv_mass}, indicates that
the background shaping is indeed largely gone.\medskip

%------------------------------------------------
\begin{figure}[t]
\includegraphics[width=0.45\textwidth]{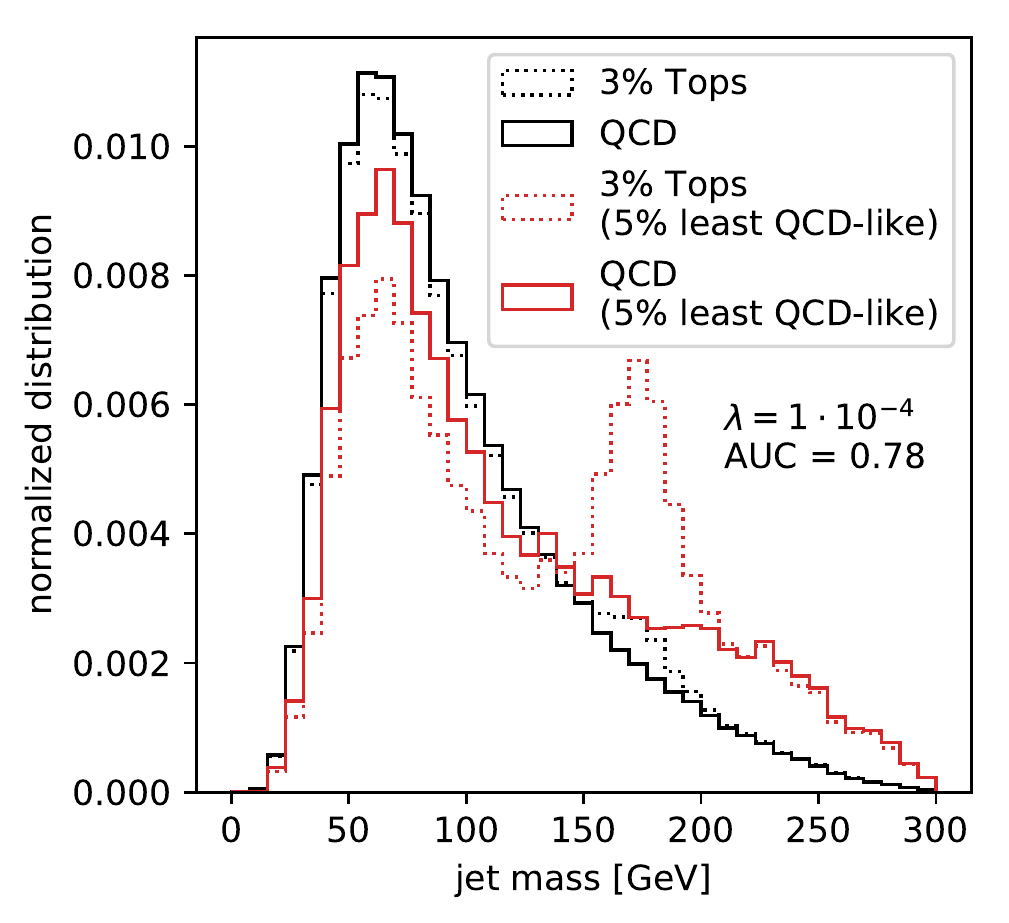}
\hspace*{0.03\textwidth}
\includegraphics[width=0.45\textwidth]{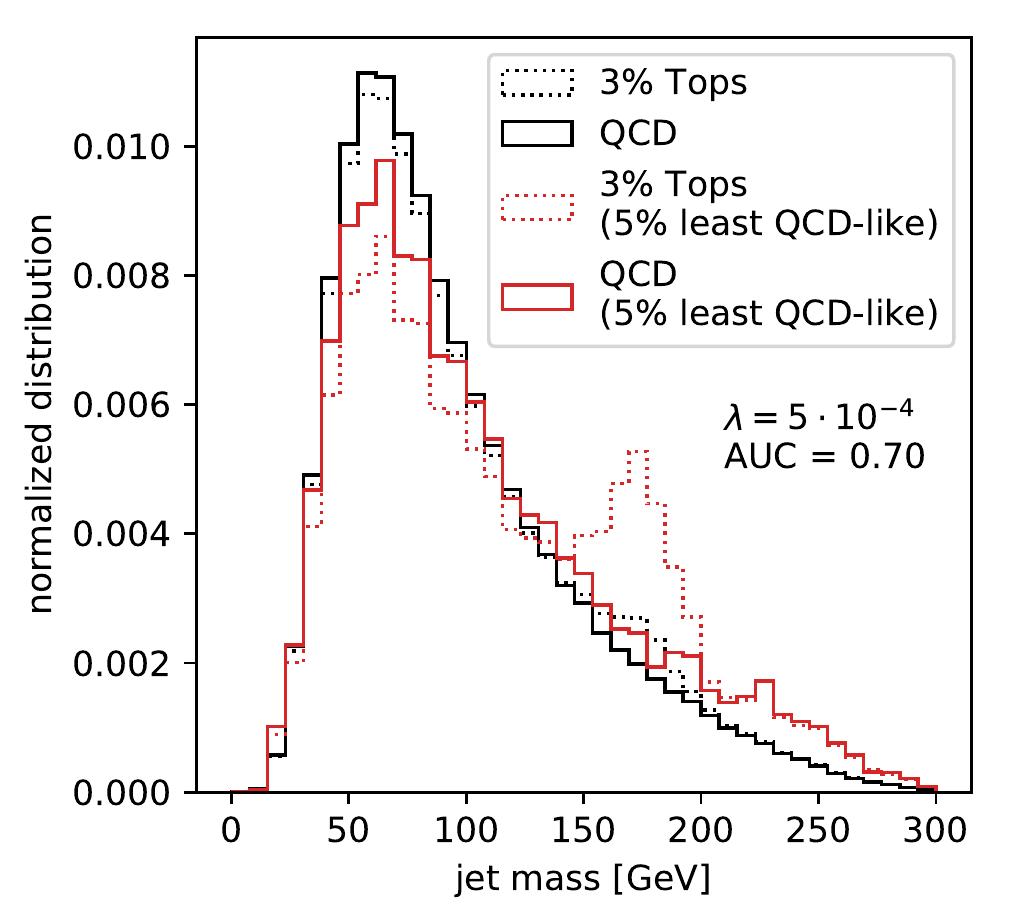}
\includegraphics[width=0.45\textwidth]{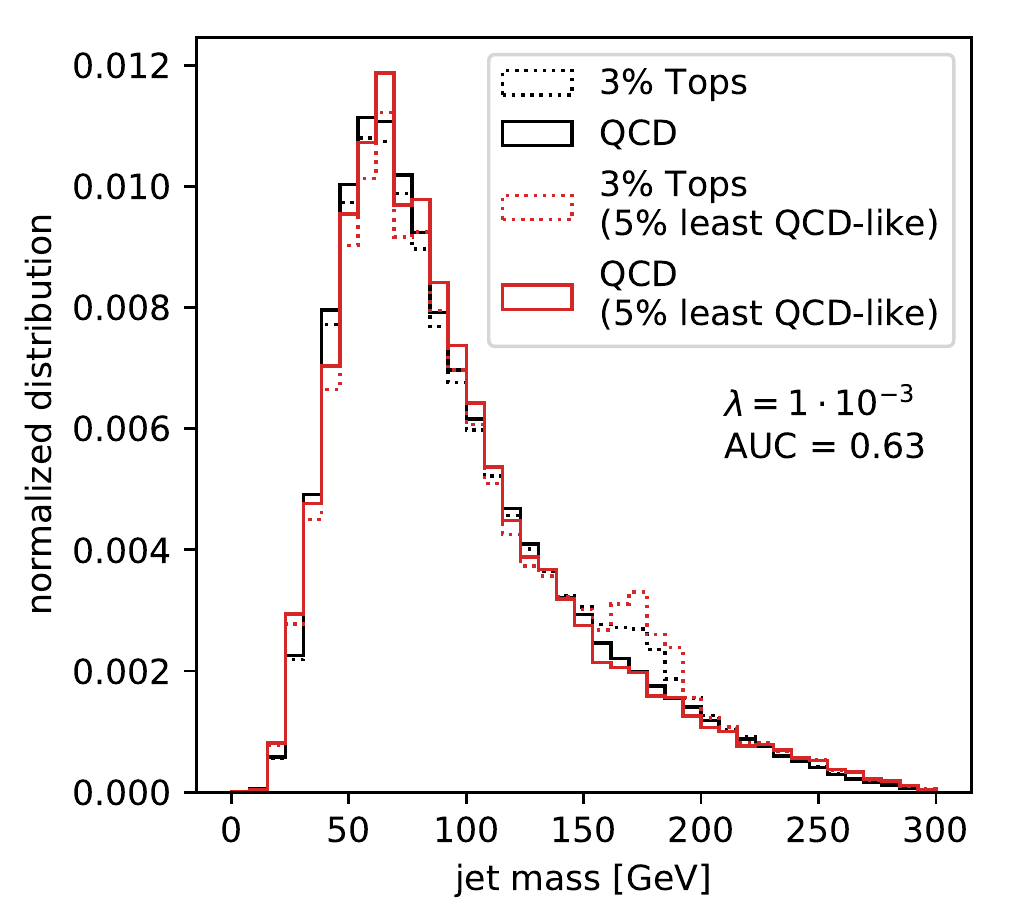}
\hspace*{0.098\textwidth}
\includegraphics[width=0.43\textwidth]{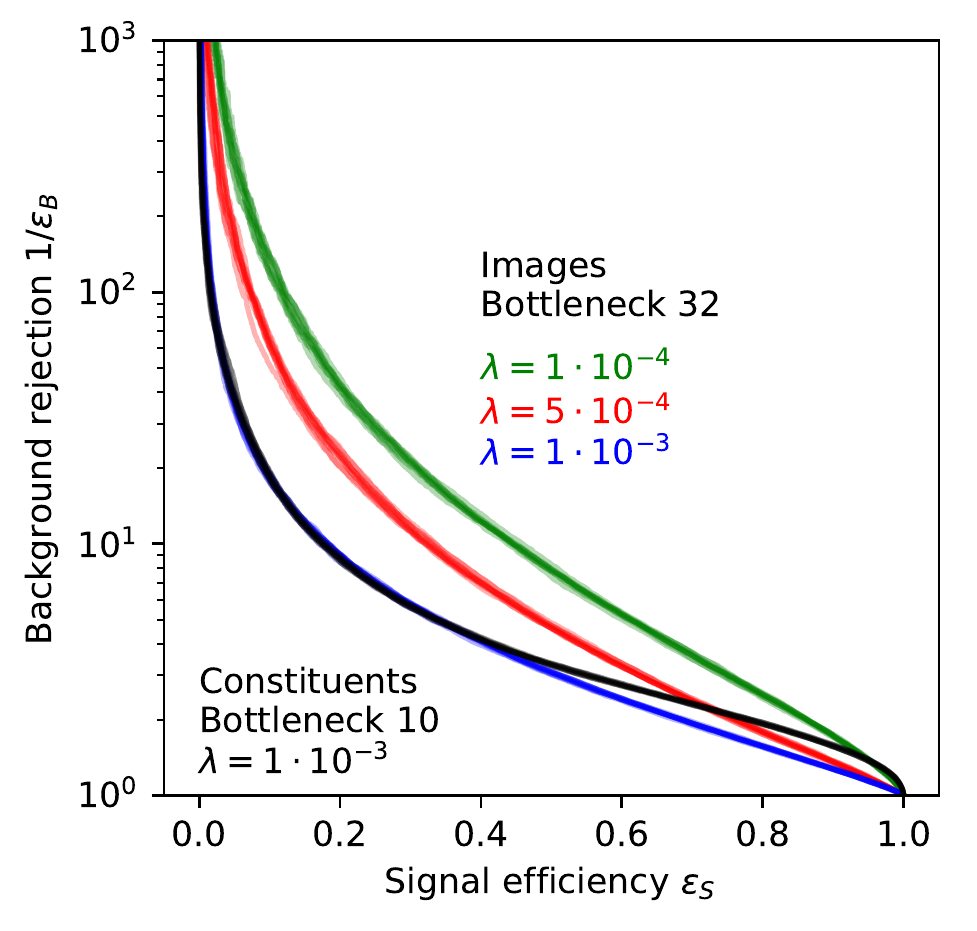}
\caption{First three panels: jet mass distributions from the
  adversarial autoencoder with different values for $\lambda$, trained
  on pure QCD, and tested on pure QCD and a sample with 3\% top
  jets. Lower right: ROC curves for the image-based and 4-vector-based
  adversarial autoencoders. The widths of the lines show the variation
  based on ten independent test samples for fixed
  training.}
\label{fig:adv_roc}
\end{figure}
%------------------------------------------------

To study the interplay of the mass de-correlation with the performance
of the adversarial autoencoder we show results for three values of
$\lambda$ in Fig.~\ref{fig:adv_roc}. For increasing values of
$\lambda$ the background shaping indeed improves. On the other hand,
we can illustrate the performance of the network by testing on QCD
data with 3\% top jets injected.  For the full sample we indeed see a
hint of top jets around $m_j = m_t$ in all three panels of
Fig.~\ref{fig:adv_roc}. We can then extract the 5\% least QCD-like
jets, which should include most of the top jets. What we find is that
the number of top jets in this selection is diluted from the maximum
expected 3/5 of the 5\% least QCD-like jets. This dilution grows with
$\lambda$, because it is an effect of taking out the jet mass as the
strongest discriminator from the network. The performance drop is
given as AUC values and detailed in the right panel of
Fig.~\ref{fig:adv_roc}, where we show the ROC curves for the
adversarial autoencoder. As before, we evaluate the network on 10
independent test samples of 20,000 QCD jets and 20,000 top jets.

For the interplay between the mass de-correlation and the performance
of the network the ROC curves are not the final word, though.  Because
the jet mass is removed from the autoencoder, we now see a clear top
mass peak in the least QCD-like selection. This peak can be extracted
using a shape analysis of the jet mass distribution with fully
controlled side bands.  This feature makes a huge experimental
difference and clearly shows how the adversary in the jet mass
promotes the autoencoder to a powerful experimental
discriminator.\medskip

Finally, we can combine the same adversary part of the network with
the 4-vector-based autoencoder described in Sec.~\ref{sec:auto_lola}.
The combined loss function is now given by Eq.~\ref{eq:loss_adv}, but
including the 4-vector-based loss function of
Eq.\eqref{eq:loss_lola}. The ROC curve for a background shaping
similar to the choice $\lambda = 5 \cdot 10^{-4}$ for the images shows
that in the \text{LoLa} setup it is much harder to de-correlate the
jet mass. Correspondingly, the networks are less stable and have a
worse performance. This is because the \textsc{LoLa} architecture in
Eq.\eqref{eq:lola} focuses the network on learning the jet mass,
which should then not be the one observable we de-correlate through
the adversarial network. For that reason, we will focus on image-based
adversarial autoencoders for the rest of this paper.

%%%%%%%%%%%%%%%%%%%%%%%%%%%%%%%%%%%%%%%%%%%%%%%%%%%%%%%%%%%%%%%%%%%%%%
\subsection{Realistic analysis setup}
\label{sec:impure}

%------------------------------------------------
\begin{figure}[t]
\centering
\includegraphics[width=0.45\textwidth]{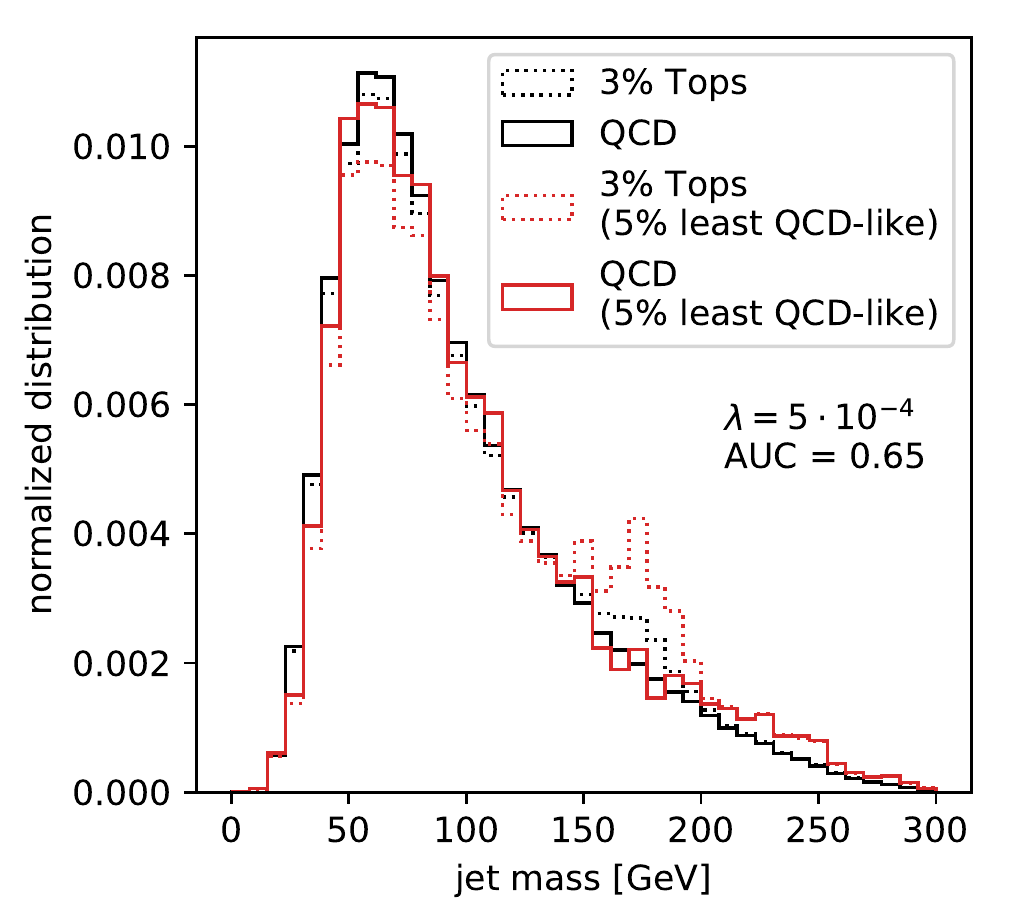}
\caption{Jet mass distributions from the adversarial autoencoder
  trained on a mixed sample with 3\% top jets using $\lambda = 5 \cdot
  10^{-4}$, and tested on a QCD sample or on the same mixed sample.}
\label{fig:impure}
\end{figure}
%------------------------------------------------

The problem in an actual analysis based on fully un-supervised learning
on QCD jets will be that we cannot avoid a certain signal
contamination of the training data. If the QCD training sample
includes a small fraction of non-QCD, or in our case top jet, the
autoencoder will accommodate top jets as QCD-like more easily.  With
the adversarial autoencoder we have developed an approach that can
identify anomalous jets uncorrelated from any variable of choice.  

In Fig.~\ref{fig:impure} we show the usual jet mass distribution for
the image-based network, but trained on a QCD sample contaminated by
3\% top jets. We keep $\lambda = 5 \cdot 10^{-4}$, but choose a much
smaller bottleneck of 10 because the network now tends to accept tops
as QCD-like jets, so we need to squeeze it harder in extracting
non-QCD features. The performance with these settings is almost the
same as for the adversarial training on QCD jets only, shown in
Fig.~\ref{fig:adv_roc}, with an AUC of 0.65 instead of 0.70. This loss
in performance can largely be recovered by a small change in the
Lagrangian multiplier to $\lambda = 3 \cdot 10^{-4}$.  Also the
remaining background shaping in Fig.~\ref{fig:impure} is similar to
the pure QCD case shown in Fig.~\ref{fig:adv_roc}. As hoped for, the
top jets in the test sample are still collected as part of the least
QCD-like jets, and they retain a distinctive mass peak which the
squeezed adversarial network does not flatten.\medskip

This behavior opens the door to new strategies searching for physics
beyond the Standard Model. We briefly sketch the application to a bump
hunt using the invariant jet mass~\cite{cwola}.  First, we define a
region of phase space and an analysis variable. While our method can
be applied to the inclusive QCD jet distribution, we focus on a search
in jets with large transverse momentum as motivated by possible signals
for hadronic decays of massive particles. Given a phase space region
we use simulated QCD jets to set the hyper-parameters of the
adversarial network, including the Lagrange multiplier. The three
figures of merit are: flatness of the mass response, ability to
identify a benchmark signal, and stability of the training.

Crucially, the actual training of the network already uses data from
the same sample as we want to analyze. This means that we split the
full data sample into statistically independent training and analysis
samples.  The key distribution is the jet mass for
increasingly anomalous jets. It can be evaluated
using standard bump-hunting techniques to extract a new physics
signal. The signal jets can then be further dissected using
orthogonal analysis techniques.

Because the training and the search rely on data in the same phase
space region, the usually leading systematics do not
enter. The remaining key uncertainty is the propensity of the network
to induce a fake bump despite adversarial training. It can be reduced
through a proper tuning of the hyper-parameters on simulation and
verified using additional control regions in data.  In case we see no
signal, the network response can be used to set exclusion limits for
arbitrary signal models. Compared to usual new physics searches the
tables are turned: instead of training the network on simulation and
applying it to data, we now train the autoencoder on data and apply it
to simulation. In turn, the related systematic uncertainties have to be
considered for exclusion limits.

%%%%%%%%%%%%%%%%%%%%%%%%%%%%%%%%%%%%%%%%%%%%%%%%%%%%%%%%%%%%%%%%%%%%%%
\section{Exotics in jets} 
\label{sec:new}

While top decay jets are a great tool to test and benchmark our
autoencoder, they are clearly not the most attractive application as
the top is a known particle. Instead, we need to show how the
autoencoder works in extracting other, exotic jets from a QCD sample
where the parameters might not a priori be known. We will rely on two
examples for this purpose: first we will test the autoencoder on a
sample which includes a Higgs-like scalar decaying to four jets. It
replaces the second, $W$-mass handle in the top jet by an increase in
the subjet multiplicity. Second, we will use a modified, dark shower
with QCD radiation as well as dark radiation off heavy dark
quarks. The dark radiation produces missing energy and modifies the
jet mass distribution, while leaving two hard jets with anomalous
radiation patterns. For both of these models we show how the
autoencoder with and without adversary can be used for a
signal-independent LHC search.

%%%%%%%%%%%%%%%%%%%%%%%%%%%%%%%%%%%%%%%%%%%%%%%%%%%%%%%%%%%%%%%%%%%%%%
\subsection{Scalar decay to jets} 
\label{sec:scalar}

As an alternative to the massive top jets we study a toy model with a
Higgs-like scalar decaying to four charm jets through two light
pseudoscalars,
\begin{align}
pp \to ( \phi \to a a \to c \bar{c} \; c \bar{c} ) \text{+jets} \; .
\end{align}
The particle masses are $m_\phi = m_t = 175$~GeV and $m_a = 4$~GeV. We
are not concerned with constraints on this toy model and choose the
scalar mass such that we can easily compare our results with the top
jet case and the pseudoscalar mass such that it decays to, for example,
charm jets.

The light pseudoscalar are will be strongly boosted, and its decays
should lead to four jets without a strong hierarchy in energy and
without a distinctive mass scale aside from the jet mass.  We simulate
the signal with \textsc{Pythia8.2.30}~\cite{pythia} and
\textsc{Delphes3.3.3}~\cite{delphes}, as usual ignoring multi-parton
interaction and pile-up.  The fat jets are anti-$k_T$
jets~\cite{anti_kt} with size $R=0.8$, defined by
\textsc{FastJet3.2.2}~\cite{fastjet} with
\begin{align} 
p_{T,j} = 475~...~525~\gev \; .
\end{align}
As before, the objects of the subjet analysis are particle flow
objects~\cite{particle_flow} from the \textsc{Delphes} E-flow. The
leptons from the charm decays are taken into account for the
calorimeter. For the pre-processing we center the jets in the
$k_T$-weighted centroid before pixelization and use a range of
$-0.75~...0.75$ for the azimuthal angle and for the rapidity.\medskip

%------------------------------------------------
\begin{figure}[t]
\includegraphics[width=0.45\textwidth]{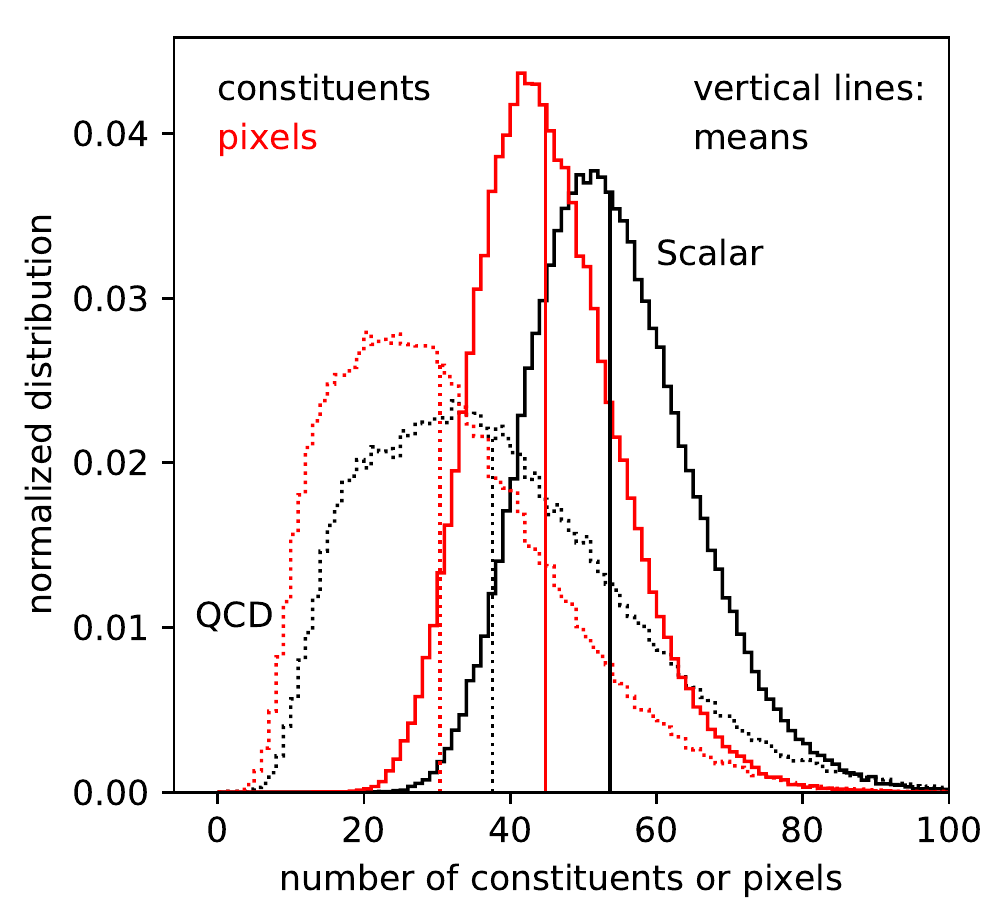}
\hspace*{0.03\textwidth}
\includegraphics[width=0.46\textwidth]{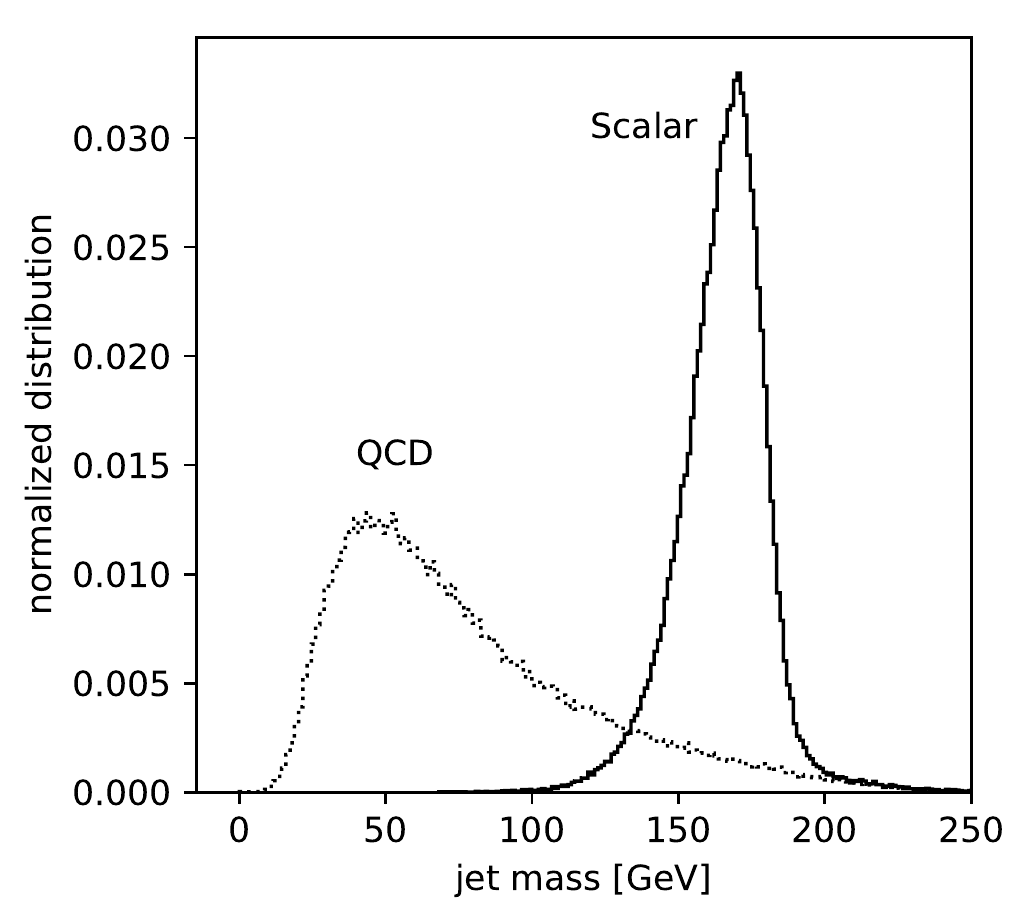}
\caption{Left: numbers of constituents and of non-zero pixels for
  scalar decay jets and QCD. Right: truth-level jet mass distributions
  for the signal and the QCD background.}
\label{fig:scalar1}
\end{figure}
%------------------------------------------------

In Fig.~\ref{fig:scalar1} we show the main physics patterns of the
scalar decay jets compared to the QCD background. In the left panel we
see the number of constituents. Comparing to Fig.~\ref{fig:roc_cnn} we
see that the general patterns are very similar, with the color-charged
top leading to a slightly larger number of constituents.  In the right
panel we show the jet masses for the signal and the background. Both
plots indicate that the heavy scalar signal is very similar to the top
signal, but without the intermediate mass drop from the
$W$-decay. This will force the de-correlated network to discriminate
signal and background just based on the number of properties of the
constituents from the scalar decay vs QCD radiation. In principle, we
could increase the reach for this model by applying $c$-tagging, but
for our toy model we explicitly do not want to use this additional
information.\medskip

%------------------------------------------------
\begin{figure}[t]
\includegraphics[width=0.43\textwidth]{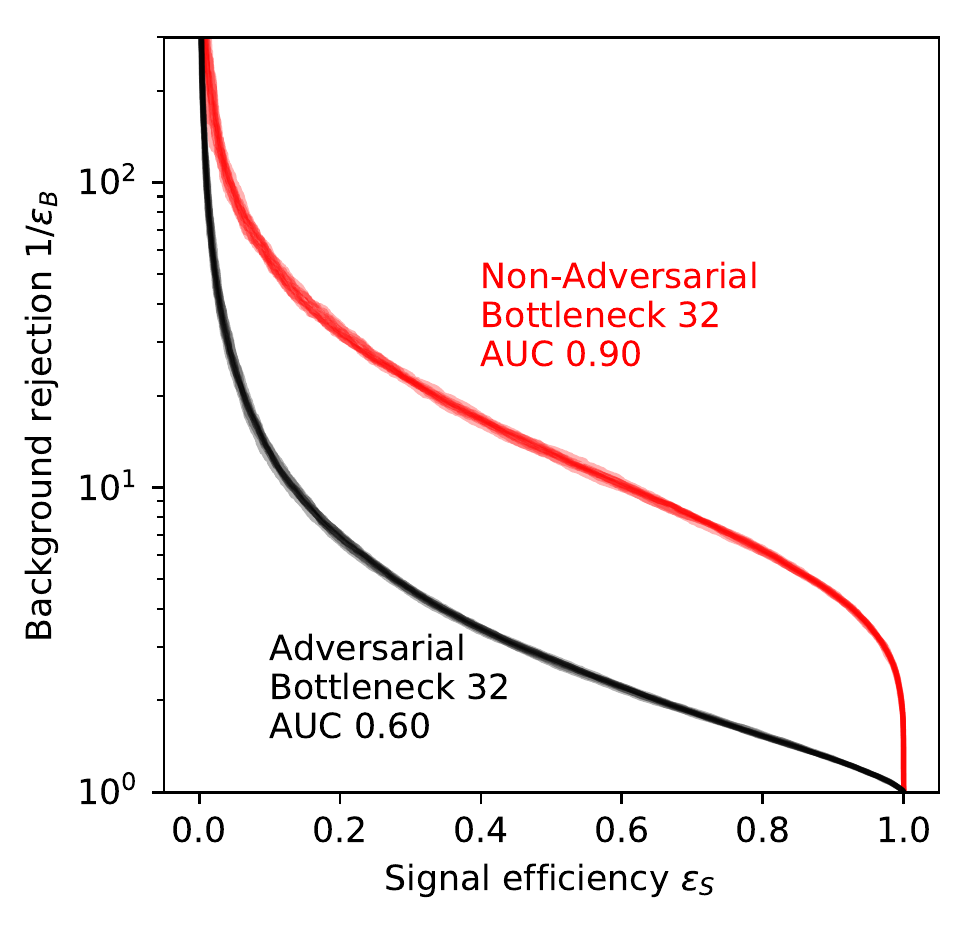}
\hspace*{0.03\textwidth}
\includegraphics[width=0.45\textwidth]{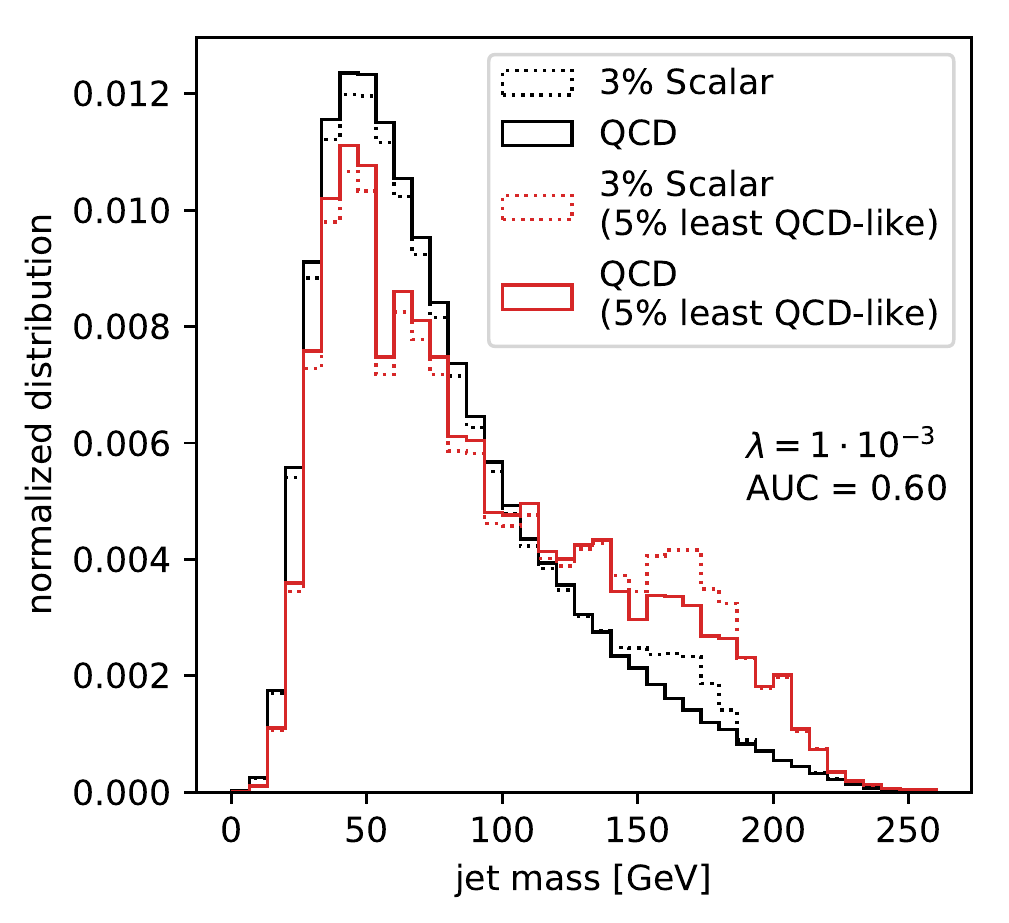}
\caption{Left: ROC curves for the image-based autoencoders with and
  without adversary. Right: jet mass distributions from the
  adversarial autoencoder trained on pure QCD.}
\label{fig:scalar2}
\end{figure}
%------------------------------------------------

The setup of the autoencoder network with and without adversary is
exactly the same as for the top case, including a bottleneck size of
32 units. The total size of the generated sample was 800,000 jets for training,
and $\approx$~250,000 jets each for validation and final testing.
In the left panel of Fig.~\ref{fig:scalar2} we include a ROC
curve for the image-based autoencoder network without adversary,
trained on QCD jets only. It corresponds to an AUC value of 0.90,
comparable to the top case. As before, we can add an adversary to the
autoencoder, to remove the information on the jet mass from the
network and to generate control samples. This leads to a weaker
performance of the network. For the same bottleneck of 32 units and a
Lagrangian multiplier $\lambda = 10^{-3}$ we find the ROC curve given
in Fig.~\ref{fig:scalar2} with an AUC value of 0.60. As mentioned
before, this is significantly worse than for the top case, because the
scalar is missing a second mass drop at intermediate masses.

To see the effect of the adversarial, we show the performance after
training on pure QCD jets and evaluated on a sample including 3\%
signal jets in analogy to Fig.~\ref{fig:adv_roc}. Two sets of curves
include all jets or the 5\% least QCD-like jets in the right panel of
Fig.~\ref{fig:scalar2}. First, we indeed observe a small enhancement
around $m_j = m_t$. While for our choice of the Lagrangian multiplier
there remains a small background shaping, we also observe a clear
signal enhancement for the least QCD-like events. However, the
scalar example also shows the limitations of a subjet analysis
where we cannot apply a mass drop and have to rely on difference
similar to quark-gluon discrimination.

%%%%%%%%%%%%%%%%%%%%%%%%%%%%%%%%%%%%%%%%%%%%%%%%%%%%%%%%%%%%%%%%%%%%%%
\subsection{Dark showers} 
\label{sec:shower}

%------------------------------------------------
\begin{figure}[t]
\includegraphics[width=0.46\textwidth]{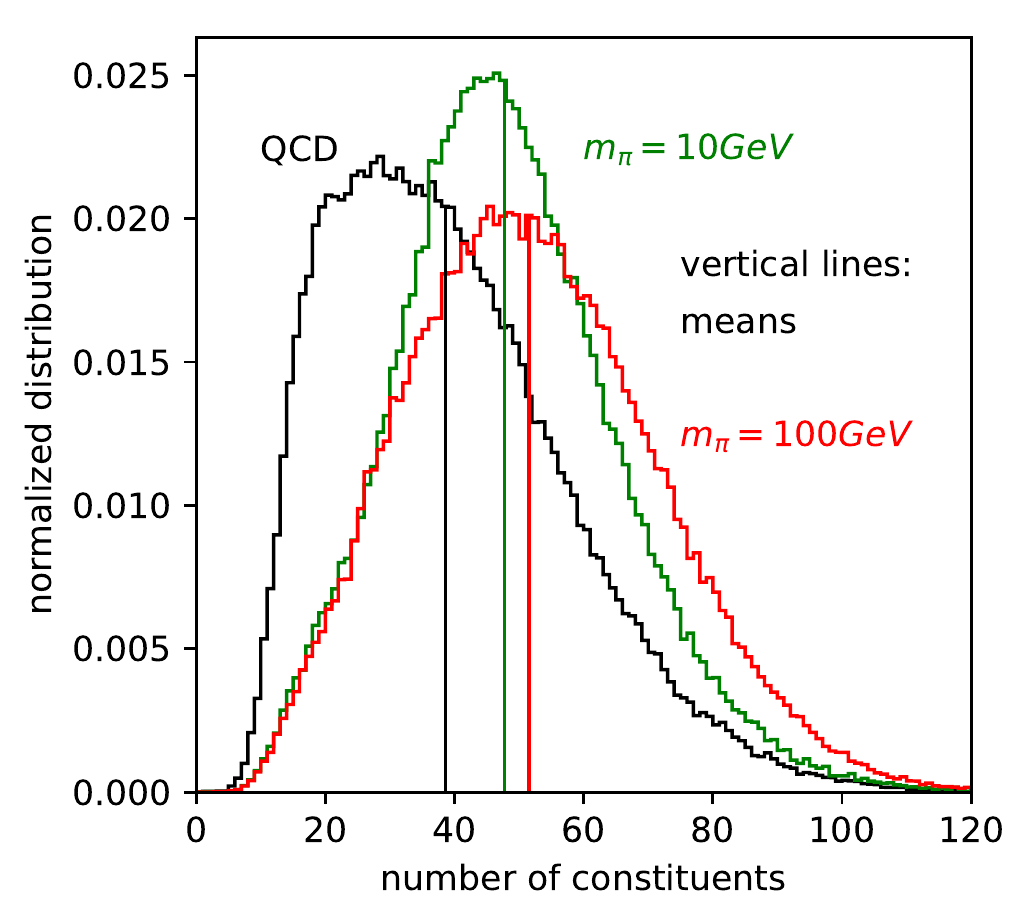}
\hspace*{0.03\textwidth}
\includegraphics[width=0.46\textwidth]{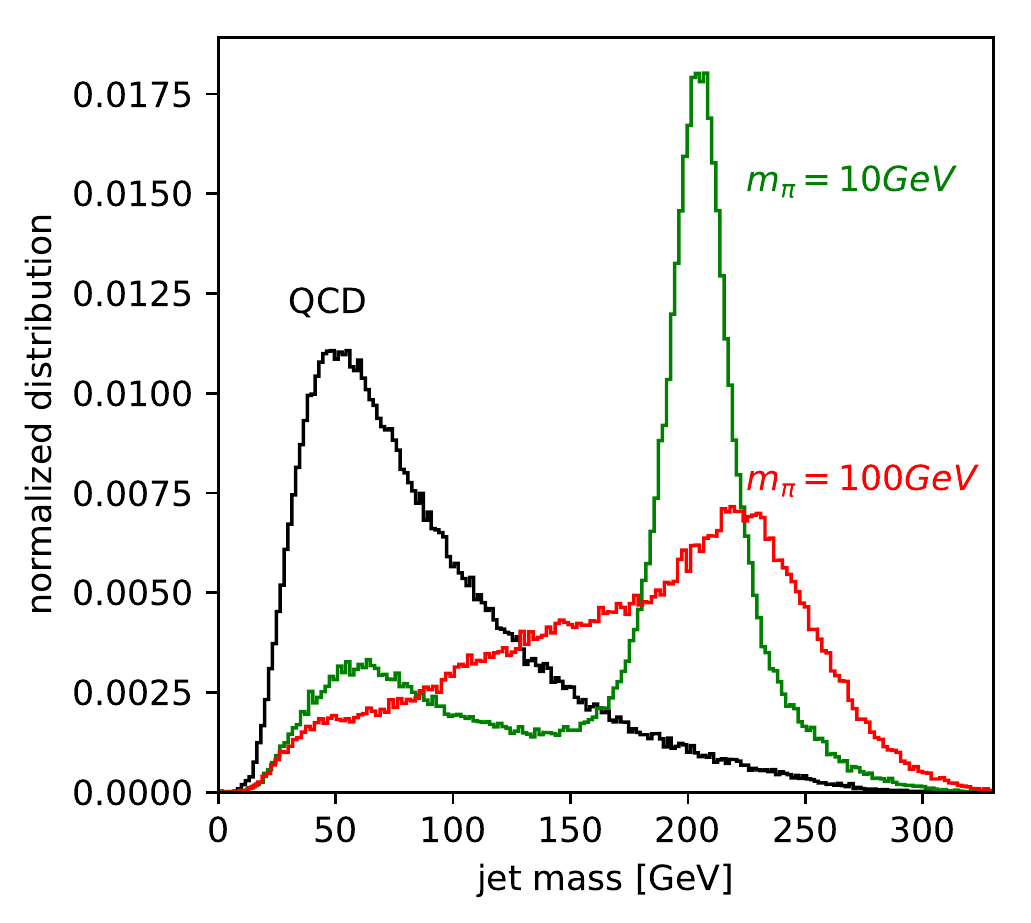} \\
\includegraphics[width=0.45\textwidth]{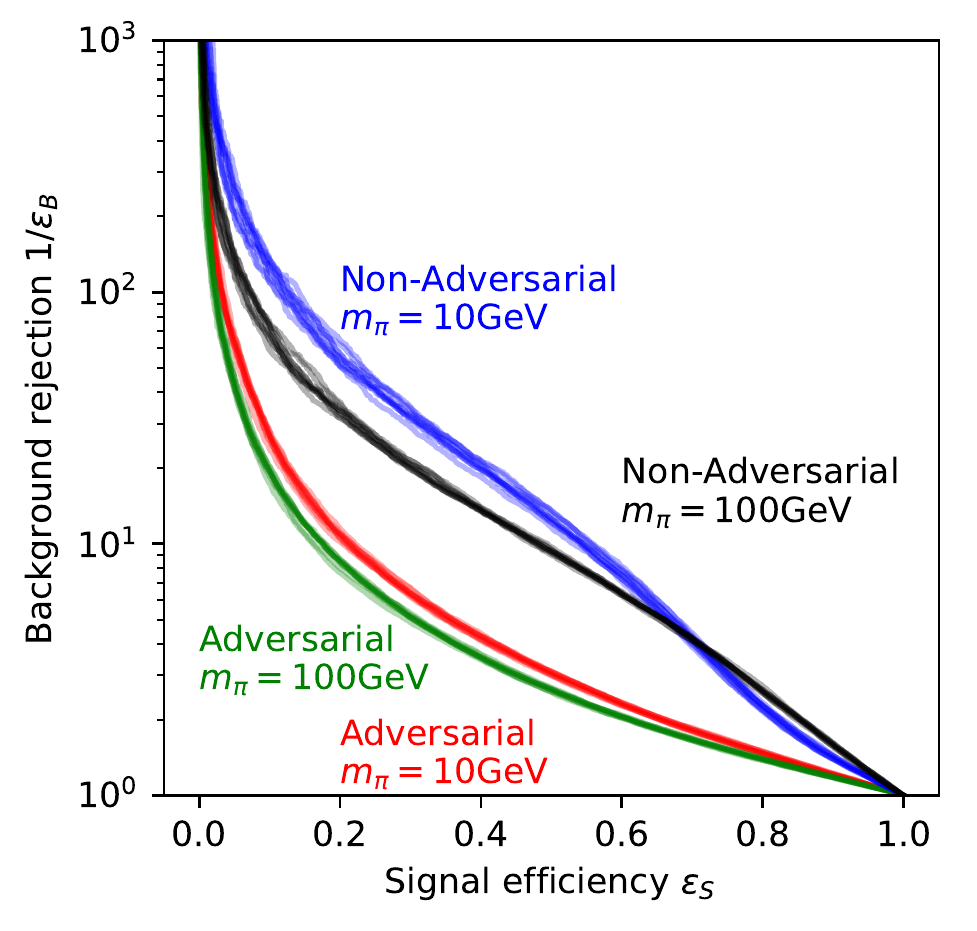}
\hspace*{0.04\textwidth}
\includegraphics[width=0.46\textwidth]{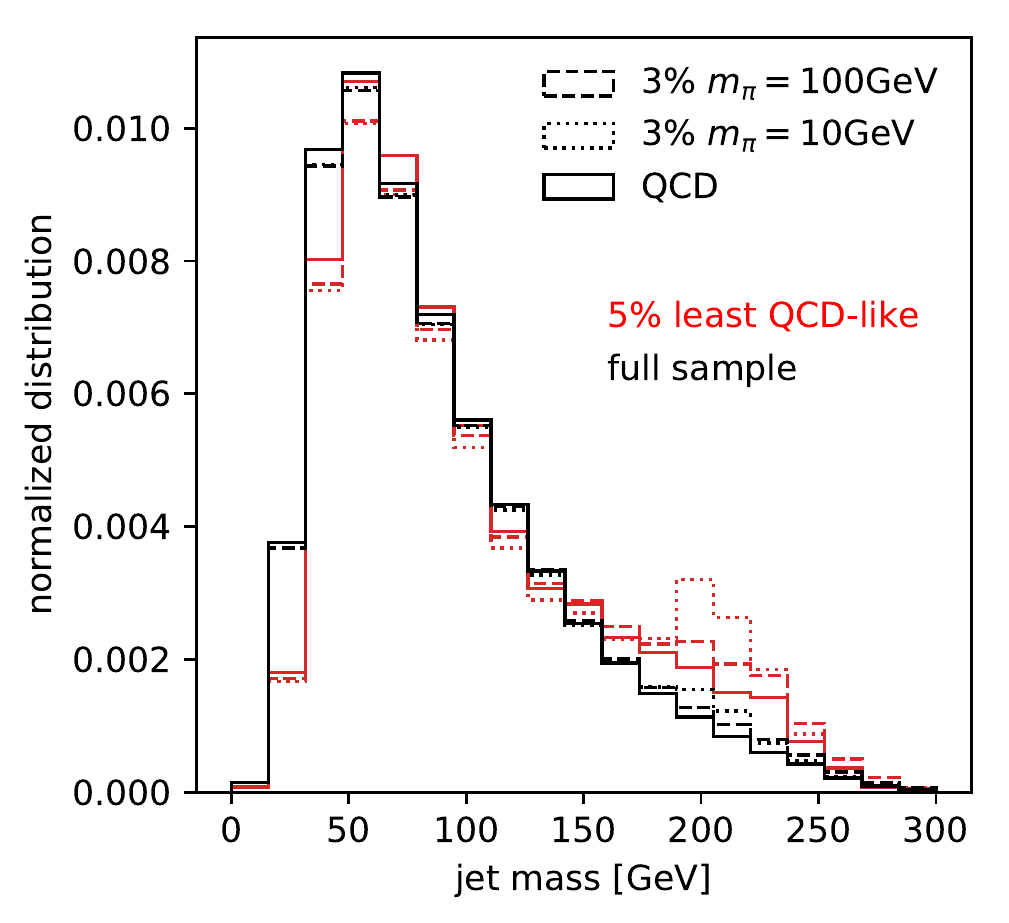}
\caption{Autoencoder applied to a set of dark shower signals. Upper
  left: numbers of constituents for the dark shower models. Upper
  right: truth-level jet mass distributions for the different models.
  Lower left: ROC curves for the autoencoders with and without
  adversary. Lower right: jet mass distributions from the adversarial
  autoencoder trained on pure QCD.}
\label{fig:shower}
\end{figure}
%------------------------------------------------

We use modified, dark showers~\cite{dark_shower}
as another benchmark scenario, independent of their 
new physics motivation through hidden valley
models~\cite{valley}. We assume that the model includes a
heavy dark quark $q_v$ which can be pair-produced at the LHC. It
undergoes showering in the dark and SM sectors and eventually decays
to its SM-partner and a light dark boson, $b_v$, which is uncharged
under all SM-gauge groups. This dark boson hadronizes into scalar and
pseudoscalar dark meson states, collectively labelled as $\pi_v$ and
assumed to have identical masses $m_{\pi_v}=2 m_{b_v}$. Depending on
the model parameters the dark mesons can decay back to SM particles
via a reverse of the production process, or leave the detector
unobserved. The visible signature is therefore di-jets plus a variable
amount of missing energy
\begin{align}
pp \to q_v \bar{q}_v
\to q \bar{q} + \met \; .
\end{align}
However, the exotic production mechanism through a heavy color-charged
dark quark leads to a sizeable amount of QCD radiation together with
the dominant jets. It generates a jet mass spectrum with an upper edge
at the dark quark mass.  For our study we use a range of dark quark
and dark boson masses.  The dark gauge interaction we consider is
$SU(3)_v$ with $\alpha_v=0.1$, which is a \textsc{Pythia} default
model.

The generation setup of for the dark showers is the same as for the
heavy scalar in Sec.~\ref{sec:scalar}, only with a slightly higher
$p_T$ range,
\begin{align} 
p_{T,j} = 575~...~625~\gev \; .
\end{align}
The image preprocessing is identical to the scalar case with 
minimal pre-processing before pixelization.\medskip

For the dark shower model parameters we again ignore current
experimental constraints and choose scenarios which best test and
illustrate the behavior of our adversarial autoencoder.  Similar to
the top and heavy scalar cases we use a dark quark with mass $m_{q_v}
= 200$~GeV. For a small meson mass of $m_{\pi_v} = 10$~GeV we see in
the upper panels of Fig.~\ref{fig:shower} that the number of
constituents and the jet mass are similar to the other new physics
scenarios in the paper. In addition, we choose a more mass-degenerate
case of $m_{q_v} = 200$~GeV and $m_{\pi_v} = 100$~GeV to test what
happens in the absence of a peak in the jet mass altogether. For the
dark shower samples we used 800000 training, 260000 validation and
280000 testing samples.

\medskip

In the lower panels of Fig.~\ref{fig:shower} we first show the
performance of the autoencoder without adversary. For both models we
find excellent performance with AUC values of 0.78~...~0.79. In the
direct comparison, the autoencoder can more easily reject the peaked
jet mass distribution, but at high efficiencies it is hard to separate
the low-mass peak from QCD. For the adversarial network with $\lambda
= 10^{-2}$ we now use 50 jet mass bins instead of the 10 used before.
We find that the performance drops to a level comparable with the
heavy scalar case with AUC value around 0.6, but with better jet mass
de-correlation. As expected, the mass peak for the 5\% least QCD-like
events is broader and less pronounced for the mass-degenerate
model. As for the scalar case, we clearly see that the autoencoder
strategy works, but also that most of the relevant information is
included in the jet mass distribution. In return, de-correlating this
key observable for background control leads to a significant drop in
performance.

%%%%%%%%%%%%%%%%%%%%%%%%%%%%%%%%%%%%%%%%%%%%%%%%%%%%%%%%%%%%%%%%%%%%%%
\section{Outlook} 

Anomalies in jets at the LHC can be extracted with the help of an
autoencoder, a neural network based on low-level data and trained on
QCD or other background samples only. We have shown that such a
network extracts boosted hadronic top decays based on jet images or
based on 4-vectors with a simplified \textsc{LoLa} structure. 
This technique is also compatible with other jet representations 
and network architectures.
Its
reduced performance as compared to specialized taggers is balanced by
reduced systematic uncertainties in the absence of a well-defined
signal model. Moreover, one autoencoder network realizing
un-supervised learning for a given phase space region can be used to
search for many different signals at the same time.

To further reduce experimental systematics, we propose to train and
use an autoencoder network in the same phase space
region. This requires full control of the background shaping. We extend
our approach to an adversarial autoencoder based on jet images,
de-correlating for example the jet mass from the training. This allows
us to sort a jet sample by the loss function describing how QCD-like
the jet is. We find (essentially) the same jet mass distribution
for each slice in the loss function. For instance top decay jets are
now collected in the least QCD-like slices and lead to a distinct peak
in the jet mass.

Next, we have shown how to train the adversarial autoencoder on data
with a signal contamination. In that case we typically make the
autoencoder more restrictive and still find that the top jets are
classified as the least QCD-like jets. We can still select them based
on the network output and search for their distinctive peak in the jet
mass distribution for non-QCD slices.

Finally, we have shown how the (adversarial) autoencoder can be used
to not only extract top decay jets, but also decays of a heavy scalar
to four quarks, or dark showers. Both of these models are
significantly harder to extract than tops at the LHC. After
de-correlating the jet mass, the different signals retain different
amounts of information, allowing us to separate them from the QCD
background. Given the universal structure of the autoencoder network
this means that the experimental LHC collaborations could make their
networks, trained on data, public and allow external groups to test if
specific models would indeed be flagged as anomalies and are hence
excluded.

\medskip
\noindent
While finishing this paper we heard of a similar, independent study,
which is published in parallel to our work~\cite{shih2}.

%------------------------------------------------
\begin{figure}[t!]
\centering
\includegraphics[width=0.9\textwidth]{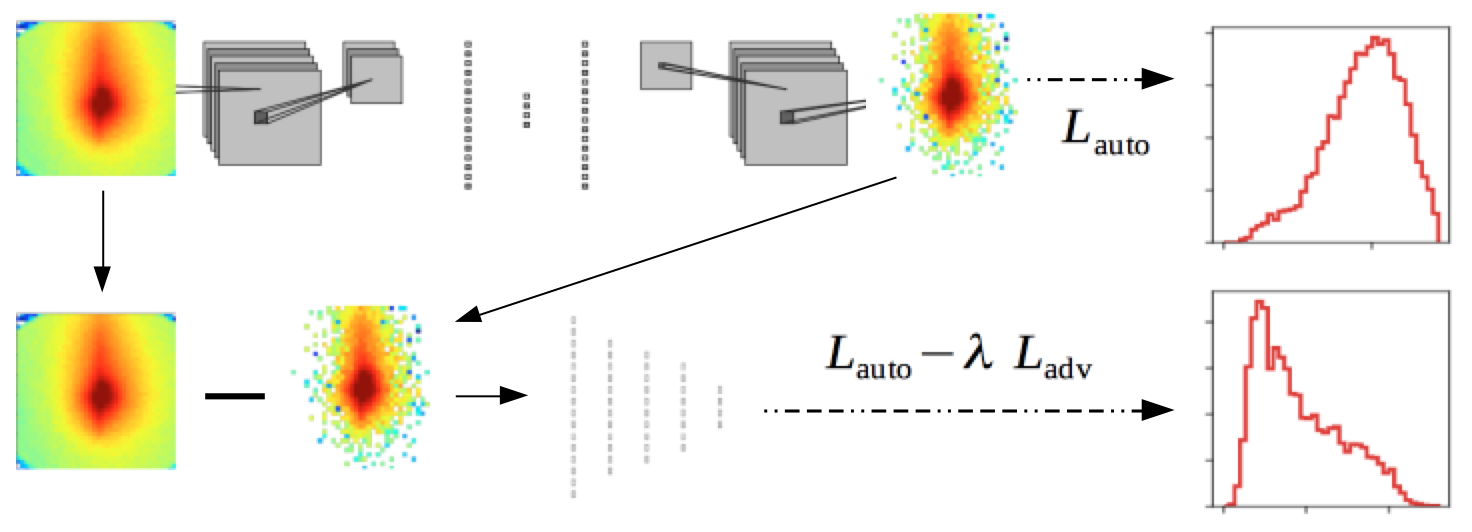}
\caption{Illustration of our network setup.}
\label{fig:illu}
\end{figure}
%------------------------------------------------

\bigskip
%%%%%%%%%%%%%%%%%%%%%%%%%%%%%%%%%%%%%%%%%%%%%%%%%%%%%%%%%%%%%%%%%%%%%%
\begin{center} \textbf{Acknowledgments} \end{center}

We would like to thank David Shih and his group for the very friendly and
constructive coordination. We are grateful to Michel Luchmann for help
with the improved image pre-processing. Finally, we would like to thank the
BOOST conference series for the encouraging atmosphere, without which papers
like this might never be written.

%%%%%%%%%%%%%%%%%%%%%%%%%%%%%%%%%%%%%%%%%%%%%%%%%%%%%%%%%%%%%%%%%%%%%%

\end{document}